\documentclass[twocolumn]{autart}    
    
\usepackage{times}
\usepackage{graphicx}        
\usepackage{multicol}        
\usepackage{amssymb,amsmath,mathrsfs}
\usepackage{mathtools}
\usepackage[inline,shortlabels]{enumitem} 
\usepackage{framed}
\usepackage{algorithm,algorithmic,psfrag}
\usepackage{soul}
\usepackage[dvipsnames]{xcolor}
\usepackage{spverbatim}
\usepackage{caption}
\usepackage[labelformat=simple]{subcaption}
\usepackage{siunitx}
\usepackage{xifthen}
\usepackage[cal=boondox]{mathalfa}  

\usepackage{tikz}
\usepackage{tikzscale}
\usepackage{textcomp}
\usetikzlibrary{spy}
\usetikzlibrary{shapes,arrows}
\usepackage{blockdiagram}
\usetikzlibrary{external}
\usetikzlibrary{positioning,calc}
\usetikzlibrary{fit}
\usepackage{pgfplots}
\pgfplotsset{compat=newest}
\pgfplotsset{plot coordinates/math parser=false}

\newtheorem{ass}{Assumption}
\newtheorem{sass}{Standing Assumption}

\newcommand{\R}{\mathbb{R}}

\newcommand{\U}{\mathcal{U}}
\newcommand{\Y}{\mathcal{Y}}
\newcommand{\zero}{\mathbf{0}}
\newcommand{\eye}{I}
\newcommand{\q}{\mathcal{q}}

\newlength\fheight 
\newlength\fwidth  

\newcommand{\DpCoeff}[1][]{%
  \ifthenelse{\isempty{#1}}%
    {\alpha}
    {\alpha_{#1}}
}
\newcommand{\DqCoeffpr}[1][]{%
  \ifthenelse{\isempty{#1}}%
    {\beta}
    {\beta_{#1}}
}
\newcommand{\DqCoeffpl}[1][]{%
  \ifthenelse{\isempty{#1}}%
    {\kappa}
    {\kappa_{#1}}
}
\newcommand{\DqCoeffql}[1][]{%
  \ifthenelse{\isempty{#1}}%
    {\gamma}
    {\gamma_{#1}}
}

\newcommand{\tcg}{\textcolor{OliveGreen}}

\newcommand{\matlab}{\mbox{\textsc{Matlab }}}
\newcommand\numeq[2]%
  {\stackrel{\scriptscriptstyle(\mkern-1.5mu#1\mkern-1.5mu)}{#2}} 

\DeclareMathOperator*{\diag}{diag}
\DeclareMathOperator*{\blkdiag}{blkdiag}

\DeclareMathOperator*{\mean}{mean}
\DeclareMathOperator*{\eig}{eig}

\begin{document}
\begin{frontmatter}
\title{Control-Oriented Modeling of Pipe Flow in Gas Processing Facilities\thanksref{footnoteinfo}} 

\thanks[footnoteinfo]{This work was supported by Solar Turbines {Incorporated}.}

\author[First]{Sven Br{\"u}ggemann}
\author[Second]{Robert H. Moroto}
\author[First]{Robert R. Bitmead}

\address[First]{Mechanical \&\ Aerospace Engineering Department, University of California, San Diego, CA 92093-0411, USA, (e-mails: \{sbruegge, rbitmead\}@eng.ucsd.edu)}
\address[Second]{R. H. Moroto was formerly with Solar Turbines {Incorporated}, San Diego CA 92123, USA (e-mail: rhmoroto@gmail.com).}

\begin{abstract}
Pipe flow models are developed with a focus on their eventual use for feedback control design at the process control level, as opposed to the unit level, in gas processing facilities. Accordingly, linearized facility-scale models are generated to describe pressures, mass flows and temperatures based on sets of nonlinear partial differential equations from fluid dynamics and thermodynamics together with constraints associated with their interconnection. As part of the treatment, the divergence of these simplified models from physics is assessed, since robustness to these errors will be an objective for the eventual control system. The approach commences with a thorough analysis of pipe flow models and then proceeds to study their automated interconnection into network models, which subsume the algebraic constraints of bond graph or standard fluid modeling. The models are validated and their errors quantified by referring them to operational data from a commercial gas compressor test facility. For linear time-invariant models, the interconnection method to generate network models is shown to coincide with automation of Mason's Gain {Formula}. These pipe network models based on engineering data are the first part of the development of general facility process control tools.
\end{abstract}
\end{frontmatter}
\section{Introduction}
Gas processing facilities, where natural gas is received, treated and compressed for onward transmission through a distribution pipeline network, provide motivation and embodiment for the development of systematic control-oriented modeling tools suited to the design of process control solutions based on plant schematics and layouts. The control of these plants involves the interconnection of a number of elements including pipes, compressors, heat exchangers, valves and valve manifolds, scrubbers and other process units and volumes. The control splits into two distinct aspects: process control for system-wide operational efficiency and accuracy, and safety systems to ensure unit and plant protection. The two control aspects differ in their timescales and in their scope, with the safety system {acting across a wide range of operating points (rather than around a single operating point)}, being {both} faster{, more highly nonlinear}, and more localized to specific unit operation, such as avoiding compressor surge. Our focus will be the process control side with an emphasis on unified plant-wide operational effectiveness. The aim of the paper is to develop interconnectable {and reconfigurable} unit system models, which are amenable to control design, with an objective of bringing multiinput-multioutput (MIMO) control into the picture for {gas processing} facilities; firstly from engineering design specifications and then augmented by data-based tuning.

\textit{Control-oriented} captures the modeling focus on eventual model-based feedback controller design reflecting: plant operational objectives, the presence and capabilities of selected actuators and sensors, and the possible reconfiguration of operations. {More precisely, our models are designed to be used for the following conditions. 
\begin{description}
\item [Plant:] Interconnected networks of {pipes and }processing elements located at one site on the order of tens of meters (rather than kilometers) in extent.
\item [Objective:] Bulk pressure regulation and disturbance flow rejection with flow {as control input/manipulated variable}.
\item [Sensing/actuation:] Sampled at or below 1Hz in line with the plant's regulation objective. The focus is on {widespread}, reliable and accurate pressure sensing in particular, and on actuation using flow {control} valves. Sensing of flow with orifice plates is there for corroboration more than for control. Temperature sensing is slow and of limited presence in the plant.
\item [Resonant and acoustic modes:] While ever-present in compression systems, are at frequencies beyond the sensor {and} actuator {bandwidth} in plants of this size.
\item [Models:] Should facilitate control design for this regime and be amenable to tuning by control-savvy plant engineers.
\end{description}
Although this is quite {a} specific scenario, it is {fairly representative} for gas processing facilities.}

{T}he models we seek will be linear(ized), time-invariant (LTI) state-space systems, optionally parametrized by nominal operating point, and capable of systematic interconnection of unit models into facility models using computer-based MIMO control design tools. {Models with time delay do not fall into this category and are therefore approximated by control-compliant dynamics if necessary.} The \textit{quid pro quo} for this {utility} is that these models are necessarily simplistic and approximate but that, by characterizing their nature, approximations might be addressed in control design. Inevitably, such modeling relies heavily on engineering knowledge of the specific application but admits fairly general applicability.

The subsystem models are based on simplified approximations to constituent equations from fluid dynamics, coupled partial differential equations (PDEs) plus algebraic equations, and are validated against plant data, including the assessment of model errors. 

Fluid dynamics and, particularly, computational fluid dynamics, are well-established subjects centered on high-fidelity modeling of flows given design and boundary conditions; typically, they involve nonlinear PDEs and transport phenomena, which are not amenable to finite-dimensional control design but instead are targeted and tested for simulation. Other pragmatic modeling for pipeline distribution systems \cite{benner2019,kralik1988,Behbahaninejad2008AMS} yields ordinary differential algebraic equations (DAEs), which again are not well suited to control design. Although, they can be used directly for controller synthesis in some circumstances \cite{mpdDae} and, as noted in \cite{benner2019}, if the DAE is of index~1. Theorem~4.1 \cite{benner2019} establishes that the DAEs are indeed of index~1 and so it is possible to rewrite the DAE as an ordinary differential equation (ODE) without the algebraic constraints. Effectively, we complete this conversion here. 

For fluid or general mechanical systems we take a lead from Benner \textit{et al.} \cite{benner2019} and Williams \textit{et al.} \cite{WilliamsKoelnPangbornAlleyneJDSMC2018} as examples where graph theoretic methods are applied to generate process models from component descriptions, with the latter paper specifically targeted at control design and the former at modeling for simulation. Williams \textit{et al.} \cite{WilliamsKoelnPangbornAlleyneJDSMC2018} is allied in its control objective with our work here and uses energy as the \textit{lingua franca} to map states between subsystems. The edges of their graphs are energy preserving connections with the dynamics occurring at the nodes. By contrast, Benner \textit{et al.} \cite{benner2019} and we model the dynamics in the edges with the nodes applying the interaction constraints. For our target processes of gas processing plants, this latter structure accords better with the primary process control objective of pressure regulation and secondarily with flow estimation. Thermal energy is a byproduct and reflection of the inefficiency of the process. While of interest, temperature is not the central manipulated variable. However, it is noteworthy that the energy formulation of Williams \textit{et al.} \cite{WilliamsKoelnPangbornAlleyneJDSMC2018} for composite aircraft systems allows conservation laws to be absorbed into the component models, so that the aggregated state-space model can be directly applied for control design. A recent comprehensive survey of modeling and feedback control design for HVAC systems is provided by Goyal \textit{et al.} \cite{GOYAL20191}, which cites Rasmussen \&\ Alleyne \cite{Rasmussen2004} who concentrate explicitly on control-oriented modeling in these vapor compression system{s}. However, because their pipes are short and well{-}insulated, the system structure again focuses on node dynamics rather than edge dynamics of our problem.

{For large-scale domain-independent systems, works from {\v S}iljak and colleagues \cite{6313113,siljakBook1978,SILJAK197275} follow a top-down approach, decomposing large-scale networks into smaller subsystems, and analyze control-relevant notions, such as (structural) stability, reachability and controllability. While their approach is generally applicable to the case of pipe flow, the logical direction differs: instead of decomposing, we \emph{compose} interconnected systems from subsystems in a bottom-up approach under the assumption that structures are fixed (rendering structural stability \cite{siljakBook1978} secondary). Further, the control actuator and, to a lesser extent, sensor locations are few when compared with the number of subsystems or network elements.}

Following \cite{benner2019}, which deals with isothermal models of gas distribution networks, we commence by studying pipe flow in individual pipes before considering how these are connected into networks yielding automatable aggregation of subsystems. The authors of \cite{benner2019} propose a network DAE with the algebraic part being the conservation of mass flows at the connection points. At this level of detail, this approach bears a strong resemblance to bond graph techniques \cite{borutzky2010bond} from which control design is problematic. However, since the resultant network DAE has index~1, the algebraic part can be solved locally to express some of the variables in terms of the others thereby eliminating them. For our models, algebraic equations arise when pipes join but not when they branch. For joints a state variable is removed yielding a new network element subsuming the three joining pipes. These new elements preserve the linearity and other properties while also respecting the conservation laws. Further, we show how these components might be aggregated into network equations to compute the larger state-space system, which we show subsumes Mason's Gain Formula. That is, we are able to preserve the simplicity of the signal flow model of the pipe network as opposed to resorting to bond graphs or DAEs.

The pipe-flow models developed are validated against {1Hz operating} data collected at the Solar Turbines {Incorporated} Gas Compressor Test Facility (GCTF) at Solar Turbines {Incorporated} in San Diego. This is a well-instrumented site normally used to test compressor performance. We use engineering design values to derive the parametrized models and then experimental data from a number of recorded tests is used to compare the fit {of} the data and model outputs. The discrepancy between model and data is used to quantify and qualify the model performance. Specifically, we find that isothermal models, such as those used in \cite{benner2019}, are subject to offsets and slow variations due to temperature gradients, which for this plant are measured but need not necessarily be. Accordingly, the control design needs to accommodate this known inaccuracy of the models. Indeed, the existing single-loop PI-controllers already give this clue and indicate that the principal plant objectives are the regulation of pressures and flows.

The design of network-ready models for pipe flow is the first stage of introducing model-based control design into these systems using engineering design information and data sheets. The project objective is to expand this to include other network elements, such as compressors, heat exchangers, vessels and valves {\cite{SRB_arXiv}, and validate their use for MIMO control design \cite{sven_bob_control}}.

\section*{Part 1: Control-Oriented Pipe Models}
We start with a deep dive into: modeling of individual pipe segments as nonlinear PDEs and boundary conditions, spatial discretization to nonlinear ODEs with input signals, then linearized ODE models with inputs. These are then compared with experimental/operational data from the GCTF, yielding control-oriented finite-dimensional linear state-space models and an appreciation of their deviation from ideal behavior. We establish that these single pipe models inherently satisfy conservation of mass flow\footnote{{This central presence of mass conservation in flow models is more fully examined in our companion model-based control design paper \cite{sven_bob_control}. There, conservation is shown to connect to integrators and inherent model structure at $s=0$, appreciation of which is critical for regulator design.}}. In Part~2, we explore how to move from pipe models to pipe network models.

\section{PDE models}
We formulate the pipe dynamics as a one-dimensional flow with standing assumptions common in the literature (e.g. \cite{ALAMIAN201251,Behbahaninejad2008AMS,benner2019,HuckTischendorf2017}). We assume these throughout the paper.

\begin{sass}\label{ass:pipeFlow} For the one-dimensional pipe flow, 
\begin{enumerate}[(i)]
\item the cross{-}sectional area of each pipe segment is constant;
\item {average velocities across the cross section suffice for the computation of the mass flow};\label{it:average}
\item there is no slip at the wall, i.e. the gas velocity at the inner pipe wall is zero; \label{it:no_slip}
\item friction along the pipe can be approximated by the Darcy-Weisbach equation, see e.g. \cite{rennels2012pipe};
\item the compressibility factor is constant along the pipe;
\item capillary, magnetic and electrical forces on the fluid are negligible.
\end{enumerate}
\end{sass}
Item \ref{it:average} is a property of high Reynolds number turbulent flow.
Under these assumptions, the constituent relations --- Continuity, Momentum, Energy, Gas Equation, respectively --- that serve as a basis for our model are 
\begin{subequations}\label{eq:constituents}
\begin{align}
&\frac{\partial \rho}{\partial t}=-\frac{\partial}{\partial x}(\rho v),\label{eq:continuity}\\
&\frac{\partial }{\partial t}(\rho v)+\frac{\partial}{\partial x}(\rho v^2+p)=-\frac{\lambda}{2d}\rho v|v|-g\rho\frac{d h}{d x},\label{eq:momentum}\\
&\q\rho =\frac{\partial}{\partial x}\left[\rho v \left(c_v T+\frac{v^2}{2}+gh+\frac{p}{\rho}\right)\right]\nonumber \\
&\qquad\qquad\qquad+\frac{\partial}{\partial t}\left[\rho \left(c_v T+\frac{v^2}{2}+gh\right)\right],\label{eq:energy}\\
&p=\rho R_sTz_0\label{eq:ideal-gas},
\end{align}
\end{subequations}
which are derived in e.g. \cite{Lurie2008} and whose parameters are defined in Table \ref{tab:parameters}. The boundary conditions \begin{align*}
p(0,t), \quad q(L,t),\quad T(0,t),
\end{align*}
are assumed to be known. Continuity Equation \eqref{eq:continuity} captures conservation of mass. Momentum Equation \eqref{eq:momentum} is obtained by a Newtonian approach considering forces acting on a fluid. {T}otal {E}nergy {E}quation \eqref{eq:energy} is the First Law of Thermodynamics in differential form, see e.g. \cite{shapiro1953vol1}. The Gas Equation \eqref{eq:ideal-gas} closely describes the behavior of natural gas at the conditions pertaining in the handling facility. 

We develop a dynamic model for $p(L,t)$ and $q(0,t)$, and if required also for $T(L,t)$, and a related methodology that allows a systematic interconnection of pipe elements in a network. Towards this goal, in Section~\ref{sec:niso3d}, from the constituent relations above we derive a nonisothermal, linear, 3D state-space model with the pressure, mass flow and temperature as state elements. Under the condition of a constant temperature, in Section~\ref{sec:isothermal2d} we revisit \eqref{eq:constituents} and introduce a simplified isothermal 2D model. In the next section we validate both models against {operating} data from the GCTF and compare them to the numerical solution of the PDEs in \eqref{eq:constituents}. This analysis suggests using the isothermal model parametrized by spatially varying nominal temperature and managing small offsets and slow drifts with the controller design. Section~\ref{sec:DAE2composite} treats the removal of algebraic constraints stemming from the DAEs and proposes a catalog of common {network units in} state-space form, including a new pipe joint element. To interconnect these unit models to pipe networks, Section~\ref{sec:connections} contains a matrix methodology, which we prove subsumes and automates Mason's Gain {Formula} in the MIMO context. The properties of interconnected components are then illustrated by a numerical experiment in Section~\ref{sec:num_ex}\footnote{{A compendium of linear state-space models for a variety of elements is provided, with derivations, in \cite{SRB_arXiv}. This paper also includes examples and \matlab\ code for interconnected networks and establishes the mass conservation property of each model.}}. We finish this paper with a brief conclusion and directions for future research.
\begin{table}[ht]
\begin{center}
\begin{tabular}{|c|l|l|}
\hline
Symbol&Meaning&SI-unit\\
\hline\hline
$A_{c}$&Cross-sectional area&$\scriptstyle[\text{m}^2]$\\
\hline
$c$&Speed of sound&$\scriptstyle[\frac{\text{m}}{\text{s}}]$\\
\hline
$c_v$&Specific heat&$\scriptstyle[\frac{\text{J}}{\text{kgK}}]$\\
\hline
$d$&Pipe inside diameter&$\scriptstyle[\text{m}]$\\
\hline
$\bar d$&Pipe outside diameter&$\scriptstyle[\text{m}]$\\
\hline
$g$&Gravity constant&$\scriptstyle[\frac{\text{m}}{\text{s}^2}]$\\
\hline
$h(x)$&Pipe elevation&$\scriptstyle[\text{m}]$\\
\hline
$k_\text{rad}$&Lumped thermal conductivity pipe&$\scriptstyle[\frac{\text{W}}{\text{m}^2\text{K}}]$\\
\hline
$L$&Pipe length&$\scriptstyle[\text{m}]$\\
\hline
$p(x,t)$&Pressure&$\scriptstyle[\frac{\text{kg}}{\text{s}^2\text{m}}]$\\
\hline
$\tilde p(x,t)$&Pressure deviation from nominal point&$\scriptstyle[\frac{\text{kg}}{\text{s}^2\text{m}}]$\\
\hline
$q(x,t)$&Mass flow&$\scriptstyle[\frac{\text{kg}}{\text{s}}]$\\
\hline
$\tilde q(x,t)$&Mass flow deviation from nominal point&$\scriptstyle[\frac{\text{kg}}{\text{s}}]$\\
\hline
$\q$&Rate of heat flow per unit area&$\scriptstyle[\frac{\text{W}}{\text{m}^2}]$\\
\hline
$Re$&Reynolds number&$\scriptstyle[1]$\\
\hline
$R_s$&Specific gas constant&$\scriptstyle[\frac{\text{m}^2}{\text{s}^2\text{K}}]$\\
\hline
$T(x,t)$&Temperature&$\scriptstyle[\text{K}]$\\
\hline
$\tilde T(x,t)$&Temperature deviation from nominal point&$\scriptstyle[\text{K}]$\\
\hline
$T_0$&Nominal temperature&$\scriptstyle[\text{K}]$\\
\hline
$T_\text{amb}$&Ambient temperature&$\scriptstyle[\text{K}]$\\
\hline
$v(x,t)$&Velocity&$\scriptstyle[\frac{\text{m}}{\text{s}}]$\\
\hline
$z$&Compressibility factor&$\scriptstyle[1]$\\
\hline
$z_0$&Constant compressibility factor&$\scriptstyle[1]$\\
\hline
$\epsilon$&Roughness of pipe wall&$\scriptstyle[\text{m}]$\\
\hline
$\lambda$&Friction factor&$\scriptstyle[1]$\\
\hline
$\rho(x,t)$&Density&$\scriptstyle[\frac{\text{kg}}{\text{m}^3}]$\\
\hline
\end{tabular}
\end{center}
\caption{Definitions of model variables and SI-units.}
\label{tab:parameters}
\end{table}

\section{Non linear and linear nonisothermal 3D ODE models}\label{sec:niso3d}
Towards a nonisothermal 3D model with pressure, mass flow and temperature as state elements, consider constituent relations \eqref{eq:constituents}. {Notice that ``3D'' refers to the number of states and not the spatial dimension.} For the corresponding total energy equation, \eqref{eq:energy}, the heat flux, $\q$, is assumed to be limited to radial conduction through the pipe, so that similar to \cite{OsiadaczChaczykowskiChemEngJ2001} and neglecting conduction through the gas,
\begin{align}
\q\rho A_{c} dx&=k_\text{rad}\pi \bar ddx (T_\text{amb}-T).\label{eq:heat-flux}
\end{align}
This characterization enables the formulation of PDEs that isolate the time derivatives of the desired state variables.
\begin{prop}\label{prop:3D_pde}
Let $|v|\ll c=\sqrt{z_0R_sT_0}$. Then, constituent equations \eqref{eq:constituents} and the heat flux described in \eqref{eq:heat-flux} yield
\begin{subequations}\label{eq:3d_pde_prop}
\begin{align}
\frac{\partial p}{\partial t} &=\frac{R_sz_0}{A_{c}c_v}\left[k_\text{rad}  \pi \bar d (T_\text{amb}-T)\right.\nonumber\\
&\quad \left.-\frac{\partial q}{\partial x}T\left(c_v+R_sz_0\right)+\frac{\partial p}{\partial x} \frac{R_sz_0 T q}{p}\right.\nonumber\\
&\qquad\qquad \left.-\frac{\partial T}{\partial x} q\left(c_v+R_sz_0\right)+\frac{\lambda R_s^2z_0^2 T^2 q^2|q|}{2dA_{c}^2 p^2}\right],\label{eq:pde_p_noniso}\\
\frac{\partial q}{\partial t}&=-A_{c}\frac{\partial p}{\partial x}-\frac{\lambda R_sTz_0}{2dA_{c}}\frac{ q\vert q\vert}{ p}-\frac{A_{c}g}{R_sTz_0} \frac{d h}{dx} p,\label{eq:pde_q_noniso}\\
\frac{\partial T}{\partial t} &=\frac{R_sz_0 T}{A_{c}c_v p}\left[k_\text{rad}  \pi \bar d (T_\text{amb}-T)-\frac{\partial q}{\partial x} TR_sz_0\right.\nonumber\\
&\quad \left.+\frac{\partial p}{\partial x} \frac{R_sz_0 T q}{ p}-\frac{\partial T}{\partial x} q\left(c_v+R_sz_0\right)\right.\nonumber\\
&\qquad\qquad\qquad\qquad\qquad \left.+\frac{\lambda R_s^2z_0^2 T^2 q^2| q|}{2dA_{c}^2 p^2}\right].\label{eq:pde_T_noniso}
\end{align}
\end{subequations}
\end{prop}
The proof is provided in the Appendix.
Proposition~\ref{prop:3D_pde} enables us to obtain a linear 3D state-space realization, through spatial discretization and subsequent linearization of these  PDEs. 

We commence with the spatial discretization using simple differences. Subscripts $\cdot_\ell$ and $\cdot_r$ connote variables at left (entry) and right (exit) sides of the pipe. Input variables are identified with the pipe PDE boundary conditions, $p_\ell$, $q_r$ and $T_\ell$, and the state variables with the ODE solution, $p_r$, $q_\ell$ and $T_r$, 
where, 
\begin{align*}
p_\ell&=p(0,t), &\quad  q_\ell&=q(0,t), &\quad  T_\ell&=T(0,t),\\
p_r&=p(L,t), &\quad q_r&=q(L,t), &\quad T_r&=T(L,t).
\end{align*}
The subscripts are motivated by the definition of a positive $x$-direction from left to right, but {do} not imply any specific flow direction, only that one is not free to prescribe both the pressure and flow at a single point. This yields the nonlinear nonisothermal 3D model:
\begin{subequations}\label{eq:3d_model_eq}
\begin{align}
\dot{p}_r&=\frac{R_sz_0}{A_{c}c_v}\left[k_\text{rad}  \pi \bar d (T_\text{amb}-T_r)\right.\nonumber\\
&\quad \left.-\frac{q_r-q_\ell}{L}T_r\left(c_v+R_sz_0\right)+\frac{p_r-p_\ell}{L} \frac{R_sz_0T_rq_r}{p_r}\right.\nonumber\\
&\quad\quad\left.-\frac{T_r-T_\ell}{L}q_r\left(c_v+R_sz_0\right)+\frac{\lambda R_s^2z_0^2T_r^2q_r^2|q_r|}{2dA_{c}^2p_r^2}\right]\label{eq:ode_p_noniso}\\
&\doteq f_p(p_\ell,p_r,q_\ell, q_r, T_\ell,T_r),\nonumber
\end{align}
\begin{align}
\dot{q}_\ell&=-A_{c}\frac{p_r-p_\ell}{L}-\frac{\lambda R_sT_\ell z_0}{2dA_{c}}\frac{q_\ell\vert q_\ell\vert}{p_\ell}-\frac{A_{c}g}{R_sT_\ell z_0} \frac{d h}{dx} p_\ell\label{eq:ode_q_noniso}\\
&\doteq f_q(p_\ell, p_r, q_\ell, T_\ell),\nonumber
\end{align}
\begin{align}
\dot{T}_r&=\frac{R_sz_0T_r}{A_{c}c_vp_r}\left[k_\text{rad}  \pi \bar d (T_\text{amb}-T_r)-\frac{q_r-q_\ell}{L}T_rR_sz_0\right.\nonumber\\
&\quad \left.+\frac{p_r-p_\ell}{L} \frac{R_sz_0T_rq_r}{p_r}-\frac{T_r-T_\ell}{L}q_r\left(c_v+R_sz_0\right)\right.\nonumber\\
&\qquad\qquad\qquad\qquad\qquad \left.+\frac{\lambda R_s^2z_0^2 T_r^2q_r^2|q_r|}{2dA_{c}^2p_r^2}\right]\label{eq:ode_T_noniso}\\
&\doteq f_T(p_\ell,p_r, q_\ell, q_r, T_\ell, T_r)\nonumber,
\end{align}
\end{subequations}
We propose the discretization from \eqref{eq:3d_pde_prop} to \eqref{eq:3d_model_eq} as it approximates reasonable well the original infinite-dimensional at low frequencies relevant for our control problem, as discussed below.

Linearizing \eqref{eq:3d_model_eq} results in the  MIMO LTI 3D state-space realization,
\begin{subequations}
\begin{align}
\dot x_t&=Ax_t+Bu_t,\\
y_t&=x_t,
\end{align}
\end{subequations}
where $A=\left.\frac{\partial f}{\partial x}\right|_{ss}, B=\left.\frac{\partial f}{\partial u}\right|_{ss}$, with $f\doteq \begin{bmatrix}
f_p& f_q&f_T
\end{bmatrix}^\top$ and $\left.\frac{\partial (\cdot)}{\partial x}\right|_{ss}$ indicating the Jacobian with respect to $x$ evaluated at steady state (denoted by subscript $ss$), and $B$ written accordingly. Further, the state and input vectors are given by the following deviations from nominal/steady-state values,
\begin{align*}
x_t=\begin{bmatrix}
\tilde p_r&\tilde q_\ell&\tilde T_r
\end{bmatrix}^\top,\quad 
u_t = \begin{bmatrix}
\tilde p_\ell&\tilde q_r&\tilde T_\ell
\end{bmatrix}^\top,
\end{align*}
{with
\begin{align*}
\tilde p_\ell&= p_\ell- p_{\ell,ss},&\quad \tilde p_r&= p_r- p_{r,ss}\\
 \tilde q_\ell&= q_\ell- q_{ss}, &\quad \tilde q_r&= q_r- q_{ss}\\
  \tilde T_\ell&= T_\ell- T_{\ell,ss},&\quad \tilde T_r&= T_r- T_{r,ss}.
\end{align*}
}
We stress that for such a state-space realization, which is the basis for modern model-based control design, the preponderance of existing tools in linear systems theory is directly applicable, such as the determination of stability, DC gains, observability and controllability. To assess sufficiency for control-oriented design, we will use this nonisothermal 3D model as a benchmark for the reduced isothermal 2D model introduced next. Where appropriate, we also compare the solution of the linear system to both the nonlinear 3D model, \eqref{eq:3d_model_eq}, and the original PDEs, \eqref{eq:3d_pde_prop}.

\section{Isothermal 2D linear ODE model}\label{sec:isothermal2d}
Assume that the temperature is constant, i.e. $T(x,t)=T_0$ for all $x\in[0,L]$ and $t\geq0$. The Continuity, Momentum and Gas Equations in \eqref{eq:constituents} suffice to obtain
\begin{subequations}\label{eq:pde}
\begin{align}
\frac{\partial p}{\partial t}&=-\frac{R_sT_0z_0}{A_{c}}\frac{\partial q}{\partial x},\label{eq:ppde}\\
\frac{\partial q}{\partial t}&=-A_{c}\frac{\partial p}{\partial x}-\frac{\lambda R_sT_0z_0}{2dA_{c}}\frac{q\vert q\vert}{p}-\frac{A_{c}g}{R_sT_0z_0} \frac{d h}{dx}p,\label{eq:qpde}
\end{align}
\end{subequations}
where for the mass flow, $q$, we additionally used the relation $q=\rho A_{c}v$. We also neglect the partial derivative of the inertia (or kinematic) term, $\rho v^2$, justified by the fact that the speed of sound, $c$, usually greatly exceeds the velocity of the fluid \cite[pp. 174]{benner2019}. This is also consistent with Proposition \ref{prop:3D_pde}.

Following \cite{benner2019}, a spatial discretization of \eqref{eq:pde} yields
\begin{subequations}\label{eq:ode}
\begin{align}
\dot{p}_r&=-\frac{R_sT_0z_0}{A_{c}L}(q_r-q_\ell),\label{eq:p_ode}\\
\dot{q}_\ell&=-\frac{A_{c}}{L}(p_r-p_\ell)-\frac{\lambda R_sT_0z_0}{2dA_{c}}\frac{q_\ell\vert q_\ell\vert}{p_\ell}-\frac{A_{c}g}{R_sT_0z_0} \frac{h}{L}p_\ell.\label{eq:q_ode}
\end{align}
\end{subequations}
Linearizing around nominal points denoted by subscript $ss$ and using tildes to denote perturbation variables, we obtain
\begin{subequations}\label{eq:linODEs}
\begin{align}
\dot{\tilde p}_r&=\DpCoeff(\tilde q_r-\tilde q_l)\label{eq:dprdt_iso}\\
\dot{\tilde q}_\ell&=
\DqCoeffpr \tilde p_r+\DqCoeffpl \tilde p_\ell+\DqCoeffql \tilde q_\ell,\label{eq:dqldt_iso}
\end{align}
\end{subequations}
with
\begin{align*}
\DpCoeff&=-\frac{R_sT_0z_0}{A_{c}L},\quad \DqCoeffpr=-\frac{A_{c}}{L},\\
 \DqCoeffpl&=\frac{A_{c}}{L}+\frac{\lambda R_sT_0z_0}{2dA_{c}}\frac{q_{ss}\vert q_{ss}\vert}{p_{\ell,ss}^2}-\frac{A_{c}gh}{R_sT_0z_0L},\\
\DqCoeffql&=-\frac{\lambda R_sT_0z_0}{dA_{c}} \frac{| q_{ss}|}{p_{\ell, ss}}.
\end{align*}
The LTI ODEs \eqref{eq:linODEs} represent a system that can be equivalently realized by
\begin{subequations}\label{eq:linearPipeModel_ss_iso}
\begin{align}
\dot x_{t} &= \begin{bmatrix}
0 & -\DpCoeff\\
\DqCoeffpr & \DqCoeffql
\end{bmatrix}x_{t}+\begin{bmatrix}
0 & \DpCoeff\\
\DqCoeffpl & 0
\end{bmatrix}u_{t},\\
y_{t} &= x_{t}
\end{align}
\end{subequations}
with $x_{t}=\begin{bmatrix}
\tilde p_{r}&\tilde q_{\ell}
\end{bmatrix}^\top$ as the state vector and $u_{t}=\begin{bmatrix}
 \tilde p_{\ell} &\tilde q_{r}
\end{bmatrix}^\top$ as the input vector.

We note immediately several properties revealed by the linear model. The elements $(\alpha,\beta,\gamma)$ of the system matrix are all negative and the matrix possesses two eigenvalues at $\frac{\gamma}{2}\pm\sqrt{\frac{\gamma^2}{4}-\alpha\beta}$. The quantity $\alpha\beta=R_sz_0T_0/L^2$ is the square of the resonant frequency of a pipe of length $L$, since $\tcg{c=}\sqrt{z_0R_sT_0}$ is the speed of sound. The friction term $\gamma$ is comparatively small. So the linearized state-space model is that of a lightly damped resonant system.

In addition to stability, the control-oriented nature of the model allows us to deduce important properties, such as controllability. {Input matrix $B$ is full row rank, so $(A,B)$ is {reachable}.} If pressure $p_r$ is measured then the system is also observable. Pressure is the simplest and most reliably measured process variable.

The DC gain from $u_t$ to $x_t$ can be readily extracted, 
\begin{align*}
G_\text{DC}=-A^{-1}B=-\frac{1}{\DpCoeff\DqCoeffpr}\begin{bmatrix}
\DpCoeff\DqCoeffpl & \DpCoeff\DqCoeffql\\
0 & -\DpCoeff \DqCoeffpr
\end{bmatrix}=\begin{bmatrix}
-\frac{\DqCoeffpl}{\DqCoeffpr} & -\frac{\DqCoeffql}{\DqCoeffpr}\\
0 & 1
\end{bmatrix},
\end{align*}
and reveals the following. In steady state:
\begin{itemize}
\item ${\tilde p_r}$ is equal to ${\tilde p_\ell}$ with appropriately signed corrections due to non-zero flow and elevation;
\item regardless of the pressure, ${\tilde q_\ell}$ is equal to $\tilde{ q_r}$ in steady state, as demanded by conservation of mass;
\end{itemize}
A more detailed analysis will be provided for the 3D state model in Section ~\ref{sec:num_ex}.

\subsubsection*{Spatial discretization}
{The spatial discretization of the PDEs using $p_\ell$ and $q_r$ as the input signals is {neither} capricious nor refractory but reflects two central matters: the boundary conditions required to specify the solution for pipe flow and the requirement for {reach}ability of the resultant state-space model. The two are not disjoint. {Assuming horizontal pipes, }the two PDEs \eqref{eq:ppde}-\eqref{eq:qpde} may be combined to yield the damped wave equation. 
\begin{align*}
\frac{\partial^2 X}{\partial x^2}-\frac{\lambda c^2}{2dA_{c}^2}\frac{q|q|}{p^2}\frac{\partial X}{\partial x}&=\frac{1}{c^2}\frac{\partial^2 X}{\partial t^2}+\frac{\lambda}{dA_{c}}\frac{q}{p}\frac{\partial X}{\partial t},
\end{align*}
for either $X(t,x)=p(t,x)\text{ or }q(t,x)$ with distinct boundary conditions. This PDE is hyperbolic and requires Dirichlet, Neumann or mixed boundary conditions at both ends to define the solutions \cite{Garabedian1986}. Pressure $p_\ell(t)$ provides the left Dirichlet boundary condition and, via \eqref{eq:qpde}, $q_r(t)$ provides the right mixed boundary condition.}

{An alternative view of this spatial discretization is that, drawing on the electrical transmission line analogue of the pipe, the voltage/pressure and current/flow at one end of the line/pipe may not be independently prescribed, since they are constrained by the driving-point impedance. From the control system perspective of this paper, the selection of $p_\ell$ and $q_\ell$ as input signals would not yield the requisite system model reachability {mentioned} above.
}

\subsection*{Cascaded pipe models}
{As discussed in the introduction, it is our primary concern to provide sufficiently accurate models for frequencies below one Hertz well-suited for process control {for} facilities with pipes of length of around tens of meters (rather than kilometers). 
Towards this goal, in Figure \ref{fig:123pipeModel} below we compare the frequency response of the a single pipe of 30m with those of two 15m pipes and three 10m pipes using the composite model for pipes in series from Section \ref{subsec:pipes_series}, which in fact represents a finer discretization. We observe that the behaviors for relevant low frequencies indeed coincide; changes for high frequencies are outside the relevant range and account for acoustical modes associated with the configurations and boundaries. Per the control objective, the bulk flow modes are preserved while the resonances fall outside the sensor and actuator bandwidths.
\begin{figure}[ht!]
    \centering
    \includegraphics[scale=.45]{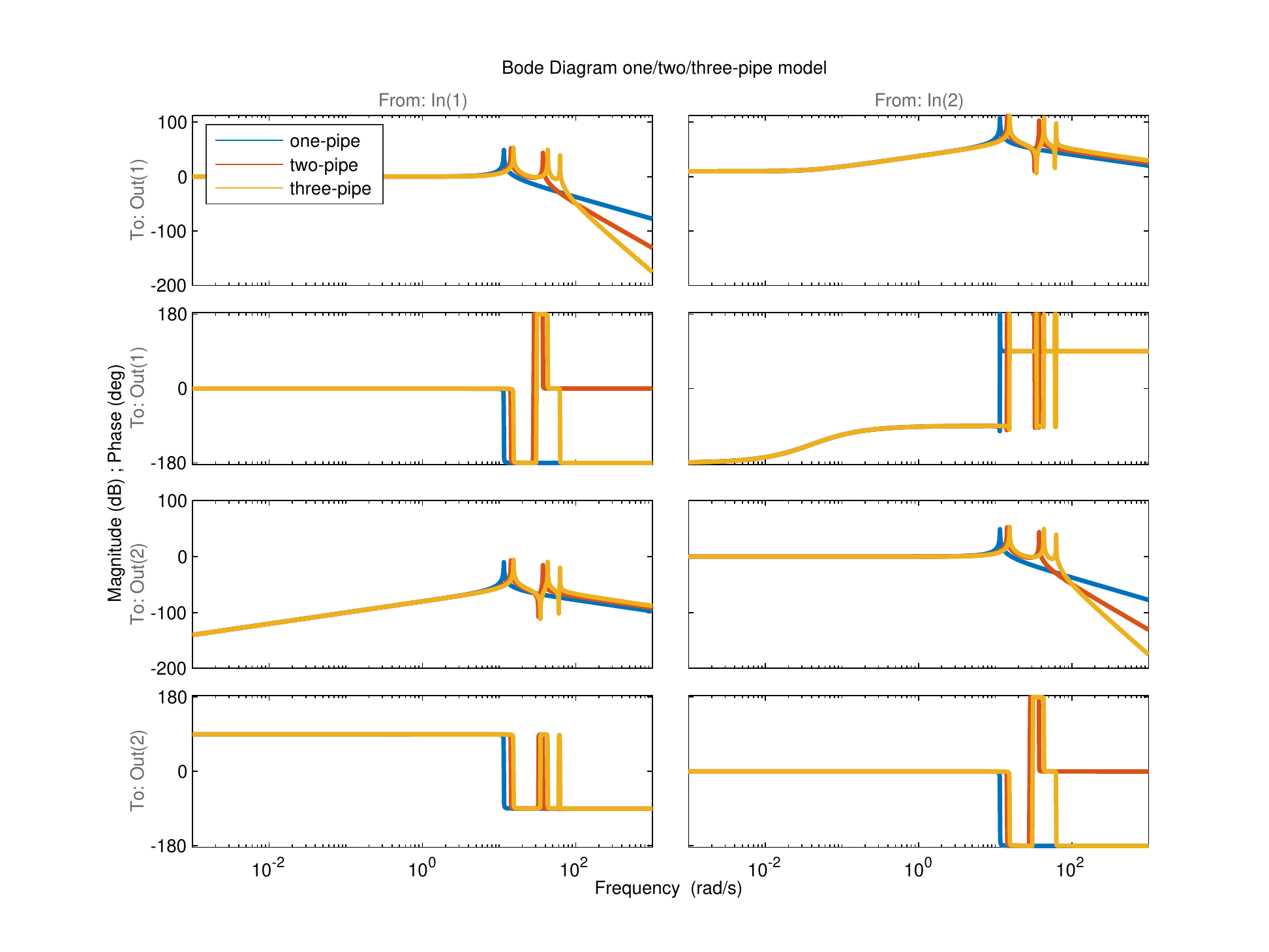}
 \caption{Comparison of the frequency responses from $\{\tilde p_{0,\ell},\tilde q_{n-1,r}\}\to\{\tilde p_{n-1,r},\tilde q_{0,\ell}\}$ with $n=\{1,2,3\}$ between one pipe ($n=1$), two pipes $(n=2)$ and three pipes $(n=3$) in series with overall identical length.}
    \label{fig:123pipeModel}
\end{figure}
} 

\subsection{Nonisothermal modeling and Bernoulli}
To ensure sufficient accuracy of linear models it is important around which nominal point they are applied. Although one may use  \eqref{eq:ode} to generate the corresponding values, we do so by solving the constituent equations in \eqref{eq:constituents} directly for steady-state values. In this fashion, firstly, we are able to accommodate spatially varying temperatures and secondly, we reveal the error inherent to the isothermal assumption and avoid its propagation.
\begin{prop}\label{prop:nonisothermal ss}
Suppose at steady state the change in density along the pipe is negligible. Then, the constituent equations in \eqref{eq:constituents} yield
\begin{subequations}\label{eq:nominal-point}
\begin{align}
q_{r,ss}&=q_{\ell,ss},\label{eq:qrss}\\
p_{r,ss}&=p_{\ell,ss}^{T_{\ell,ss}/ T_{r,ss}} \exp\left(\frac{\lambda Lz_0R_sT_{r,ss}}{2dA_{c}^2p_{r,ss}^2}q_{r,ss}|q_{r,ss}|\right.\nonumber\\
&\qquad\qquad\qquad\qquad\qquad\qquad\left.-\frac{gh}{R_sz_0T_{r,ss}}\right).\label{eq:prss}
\end{align}
If further $|v|, |h|\ll c$, $d\geq \frac{\lambda}{2}$, and $L|v|\ll c$, then
\begin{align}
p_{r,ss}&\approx p_{\ell,ss}^{T_{\ell,ss}/T_{r,ss}}\left(1-\frac{\lambda Lz_0R_sT_{r,ss}}{2dA_{c}^2p_{r,ss}^2}q_{r,ss}|q_{r,ss}|\right.\nonumber\\
&\qquad \qquad \qquad \qquad\qquad \qquad\left.-\frac{gh}{R_sz_0T_{r,ss}}\right),\label{eq:prss_aproxx}
\end{align}
where $T_{\ell,ss}=T(0)$ and $T_{r,ss}=T(L)$ at steady state.
\end{subequations}
\end{prop}
\begin{pf} For brevity, we drop subscript $ss$ in this proof.
The nominal mass flow in \eqref{eq:qrss} follows directly from the continuity equation \eqref{eq:continuity} by setting the time derivative to zero.

For the nominal pressure in \eqref{eq:momentum}, for the left-hand side, Lurie shows in \cite{Lurie2008} that 
\begin{align*}
\frac{\partial }{\partial t}(\rho v)+\frac{\partial}{\partial x}(\rho v^2)=\rho\left(\frac{\partial v}{\partial t}+v\frac{\partial v}{\partial x}\right).
\end{align*}
Now assume we are at steady state, so that for $\frac{\partial v}{\partial t}=0$ and \eqref{eq:momentum}, 
\begin{align}
v\frac{\partial v}{\partial x}dx&=-\frac{1}{\rho}\frac{\partial  p}{\partial x}dx-\frac{\lambda}{2d}v|v|dx-gdh,\nonumber\\
\frac{1}{g} vdv&=-\frac{1}{g\rho}dp-\frac{\lambda}{2dg}v|v|dx-dh,\label{eq:to_bernoulli}
\end{align}
with length $dx$. We used the fact that the change in velocity, $dv$, and pressure, $dp$, along a control volume at steady state is exactly $\frac{\partial(\cdot)}{\partial x}dx$.
Without loss of generality we now assume that the height at $x=0$ is zero. Additionally, under the hypothesis and \eqref{eq:qrss}, we can treat the velocity as a constant so that integrating \eqref{eq:to_bernoulli} along the pipe using \eqref{eq:ideal-gas} yields
\begin{align}\label{eq:bernoulli}
\frac{v_r^2-v_\ell^2}{2g}&=-\frac{R_sz_0}{g}\left(T_r \ln p_r-T_\ell\ln p_\ell\right)-\frac{\lambda L}{2dg}v|v|-h.
\end{align}
As $v_r=v_\ell$ and towards an expression for $p_r$, 
\begin{align*}
0&=-\ln\left( \frac{p_r^{T_r}}{p_\ell^{T_\ell}} \right)-\frac{\lambda L}{2dR_sz_0}v_r|v_r|-\frac{gh}{R_sz_0},\\
p_{r}&=p_\ell^{T_\ell/T_r} \exp\left(-\frac{\lambda L}{2dR_sz_0T_r}v_r|v_r|-\frac{gh}{R_sz_0T_r}\right),
\end{align*}
which with \eqref{eq:ideal-gas} gives \eqref{eq:prss}.
By the additional hypothesis, 
\begin{align*}
|hg|&\ll c^2=RsT_rz_0,\\
\lambda\frac{L}{2d}v_r^2&\leq Lv_r^2\ll c^2 = RsT_rz_0,
\end{align*} so that
\begin{align*}
p_{r}&\approx p_\ell^{T_\ell/T_r}\left(1-\frac{\lambda L}{2dR_sz_0T_r}v_r|v_r|-\frac{gh}{R_sz_0T_r}\right)\\
&=p_\ell^{T_\ell/T_r}\left(1-\frac{\lambda Lz_0R_sT_r}{2dA_{c}^2p_r^2} q_r|q_r|-\frac{gh}{R_sz_0T_r}\right),
\end{align*}
using again the {G}as {E}quation, \eqref{eq:ideal-gas}.
\end{pf}

\subsubsection*{On the assumptions}
For better understanding of conditions under which the assumptions hold  and to underline the model's suitability for control, consider Methane with $R_s=518.28\frac{\text{J}}{^\circ\text{K mol}}$, a low temperature of $\tilde T_r=300^\circ$K and a constant compressibility factor $z_0=0.95$. The related speed of sound within the medium is $c=14.77\times 10^4\frac{\text{m}}{\text{s}}$. Hence, the assumptions on the gas velocity, $v$, height, $h$, and length, $L$, conform to typical values in our control domain of gas processing facilities. Also, given a usual friction factor $\lambda\ll 1$, the lower bound on the diameter, $d$, renders our formula applicable to many industrial scenarios.

\subsubsection*{Relation to Bernoulli's Equation and isothermal model}
The proof of Proposition \ref{prop:nonisothermal ss} is of interest in itself since it delineates the relation between the dynamic Momentum {E}quation, \eqref{eq:momentum}, and static Bernoulli's Equation, \eqref{eq:bernoulli}, commonly used for computing static variables,  including a term for head loss, $H_L\doteq\frac{\lambda L}{2dg}v|v|$, often referred to as the Darcy-Weissbach Equation \cite{rennels2012pipe}. Furthermore, observe that the approximated nominal point, \eqref{eq:prss_aproxx}, coincides with the nominal point derived by  the discretized model, \eqref{eq:ode}, under the isothermal assumption and negligible change in density. In other words, Proposition \ref{prop:nonisothermal ss} also quantifies the error induced through the isothermal assumption.

\section{Model validation}\label{sec:model_validation}
We now wish to assess both the isothermal and nonisothermal models in light of their suitability for control-oriented design, using operational industrial process data from the GCTF.
{The data fits the problem formulation: it is sampled at 1Hz and describes pressure, mass flow and temperature variations for pipes on the order of tens of meters. Accordingly, it is {appropriate} for model validation and tuning {for this application}.}
Our conclusion is that, for pipe component modeling, the isothermal 2D model is sufficient for model-based control because: temperature variations in these elements are modest, temperature sensing devices can be both limited in number and variable in dynamic response, and variations with temperature can be accommodated by an appropriate controller since they are slowly varying and cause quantifiable gain fluctuations.

\begin{figure}[b]
    \centering
   \resizebox{\columnwidth}{!}{
   \includegraphics{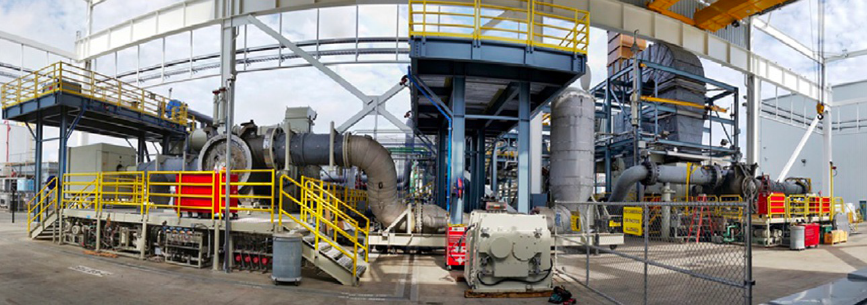}}
  \caption{Gas compressor test facility (GCTF) at Solar Turbines {Incorporated}. \cite{DELGADOGARIBAY2019449}}
    \label{fig:gctf}
\end{figure}

\begin{figure}[t]
    \centering
    \includegraphics[width=\columnwidth]{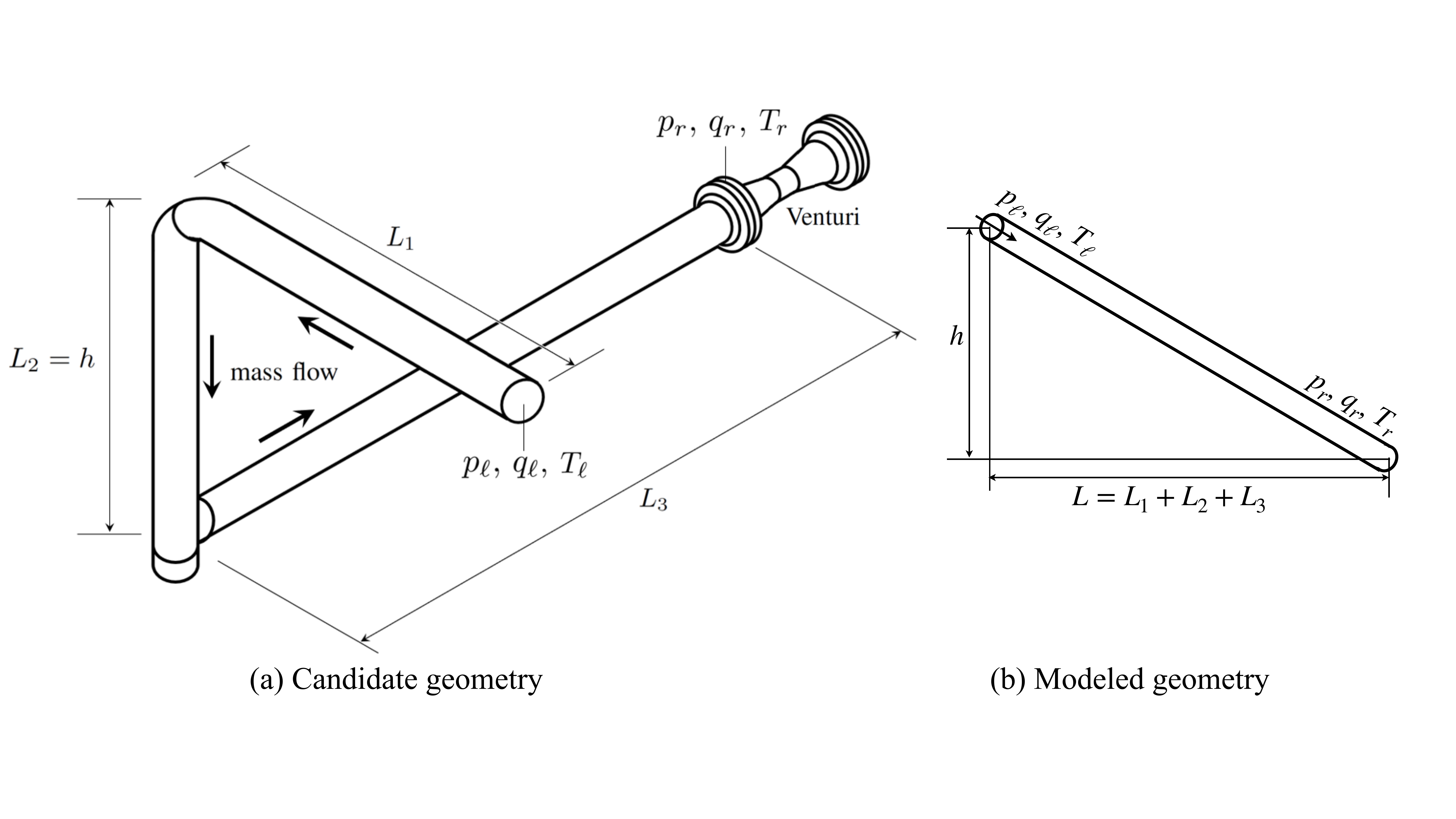}
    \caption{GCTF pipe section considered for model validation.}
    \label{fig:gctf_sketch}
\end{figure}

\begin{figure}[ht]
    \centering
\includegraphics[width=\columnwidth]{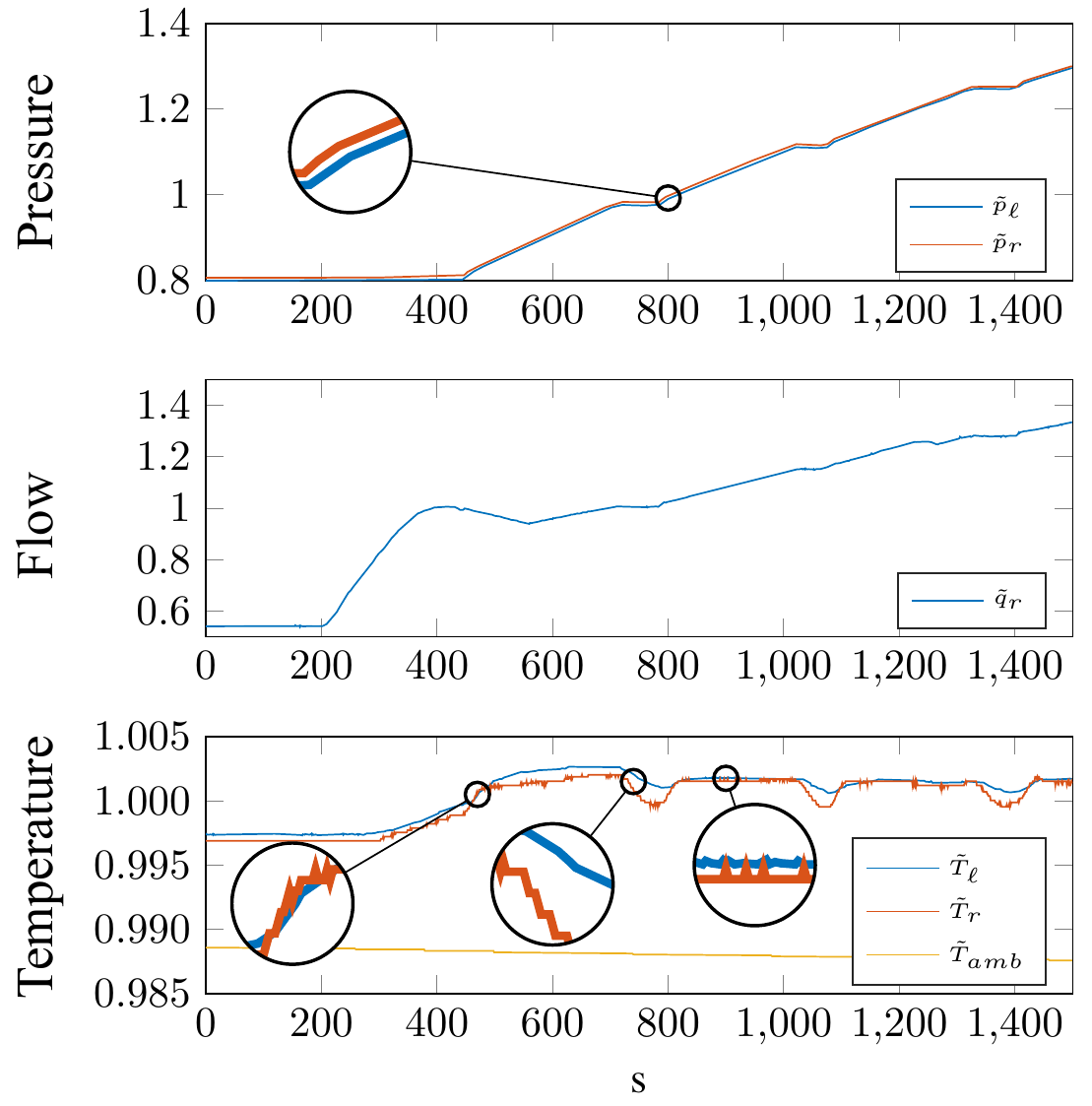}
 \caption{Normalized data from GCTF. The variables $\tilde p_\ell$, $\tilde q_r$ and $\tilde T_\ell$ will be used as model inputs, whereas $\tilde p_r$ and $\tilde T_r$ will be used to validate the corresponding model outputs. {We observe quantization errors and measurement noise.}}
    \label{fig:dataplots}
\end{figure}

Figure~\ref{fig:gctf} shows the facility at Solar Turbines {Incorporated}. This is a well{-}instrumented site used for compressor testing and from which comprehensive data sets are available.
The particular pipe section under consideration is sketched in Figure \ref{fig:gctf_sketch}.
Notice that we simplify the stepped pipe geometry by neglecting the stub at the end of the vertical middle section, assuming instead a constant slope and an accordingly adjusted friction factor\footnote{We use Haaland's formula \cite{Haaland1983} to estimate the friction factor for the straight pipes and empirical formulas in \cite[Ch. 15]{rennels2012pipe} to approximate the friction losses induced by the bends and stub.}. The relevant data is plotted in Figure \ref{fig:dataplots}, {with behavior in the relevant time scale for our goal of relatively slow process control.} We observe that the output pressure, $\tilde p_r$, closely follows the input pressure, $\tilde p_\ell$, and is higher despite head losses through friction. 
This is due to the vertical middle section and heat flux causing changes in temperature. As Figure \ref{fig:dataplots} shows, the temperature is relatively constant, but varies with changes in pressure and mass flow. We notice that $\tilde T_\ell$ is measured by a more accurate sensor, given its lower quantization error, which can be observed in the middle zoom. We additionally note that $\tilde T_r$ is lower than $\tilde T_\ell$ for most of the time, as is also apparent in the zoom on the right, and it seems to be dynamically faster than $\tilde T_\ell$, as shown in the left zoom, which may partly be caused by different thermal inertias and processing of the sensors. Given the speed, accuracy and prevalence of pressure sensors, it is apparent that they will provide the primary signals used for feedback control and the quality of capturing the pressure state behavior should be the main model objective. We shall return to this shortly in Subsection~\ref{subsec:isoNiso}.

\subsection{Linear nonisothermal 3D model}\label{subsec:lin_noiniso_mod}
\begin{figure}[t]
    \centering
    \includegraphics[width=.95\columnwidth]{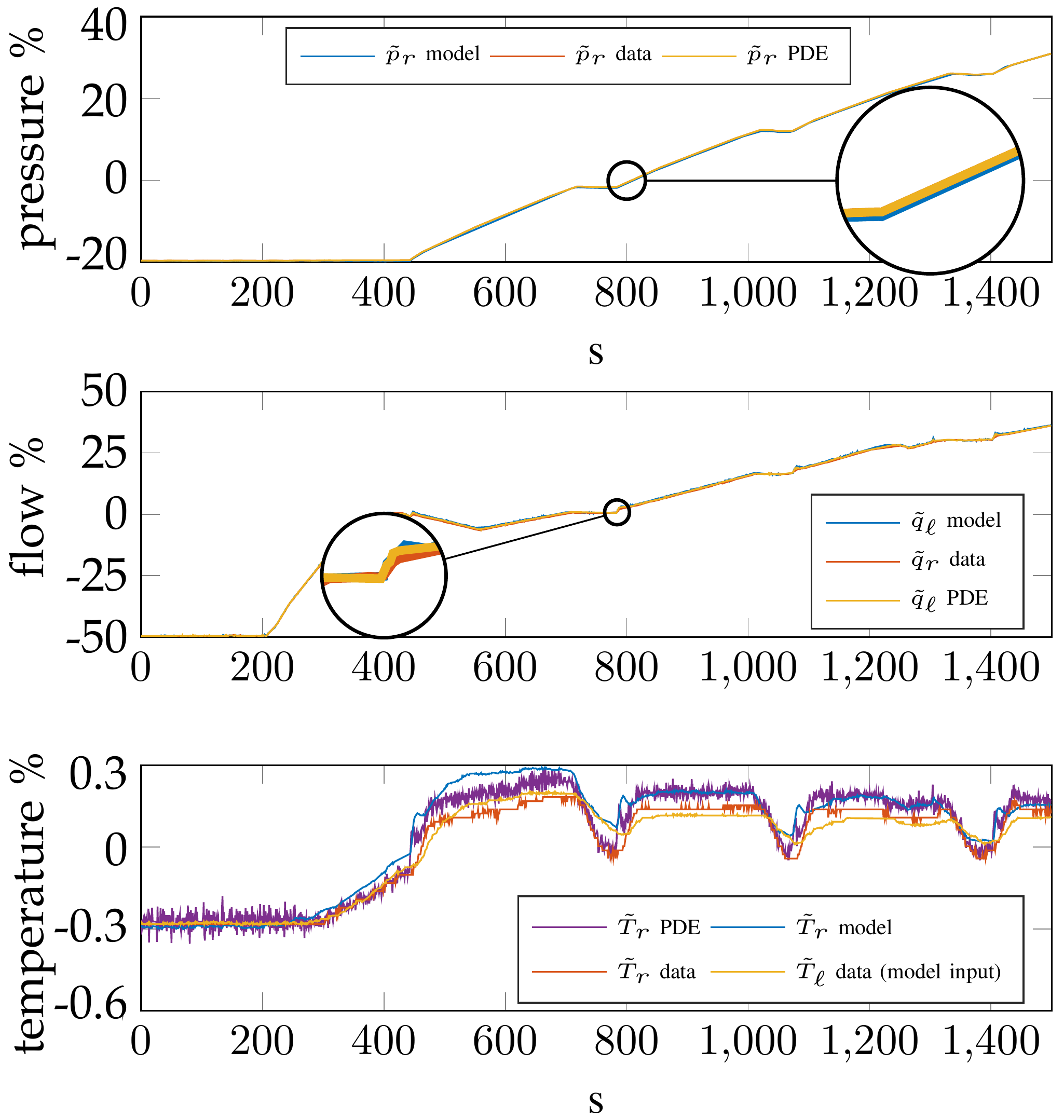}
 \caption{Percentage deviations from the nominal point {(determined as shown above)} of the nonisothermal 3D linear model with data as model inputs, compared against GCTF data and PDE model  \eqref{eq:constituents}.}
    \label{fig:3d_vs_data}
\end{figure}

We begin by validating the nonisothermal 3D model from Section \ref{sec:isothermal2d} linearized around the nonisothermal nominal point developed in Proposition \ref{prop:nonisothermal ss}. In particular, let $q_{r,ss}=\mean (q_{r}(t)),$ $p_{\ell,ss}=\mean (p_\ell(t)),$ $T_{\ell,ss}=\mean (T_\ell(t))$ and $T_{r,ss}=\mean (T_{r}(t))$ so that Proposition \ref{prop:nonisothermal ss} yields the corresponding nominal values $p_{r,ss}$ and $q_{\ell,ss}$. We then use $\tilde p_\ell$ and $\tilde q_r$ from the data set as model inputs and compare our modeled pressure $\tilde p_r$ against the related pressure in the data, recalling that data of $\tilde q_\ell$ for comparison with our model output is not available. Additionally, we study the linear model against the PDEs in \eqref{eq:constituents} solved numerically as a two-point boundary value problem, and to which we refer as the \emph{PDE model}. The lumped thermal conductivity, $k_\text{rad}$, is approximated at quasi steady state following \cite[Section 3]{OsiadaczChaczykowskiChemEngJ2001}, using $T_{\ell,ss}$ and $T_{r,ss}$. The simulation results are shown in Figure~\ref{fig:3d_vs_data}. 

For the nonisothermal 3D linear model, we see that the simulated output $\tilde p_r$ is in a small neighborhood of the measured pressure, but has a small offset. Additionally, we observe that the modeled mass flow $\tilde q_\ell$ is close to the data input, $\tilde q_r$. This is expected for the short length of the pipe $L\approx30$m and sampling rate of once per second. The zoom reveals that when the pressure increases at around 800s, the model output $ \tilde q_\ell$ first increases before the signal input $ \tilde q_r$ follows suit. This is consistent with a positive mass flow that increases first at the gas entry side of the pipe.
%
%

The temperature calculations from the successive models, while close (within $0.64$K), exhibit more variability than those of pressure and mass flow. The computed $\tilde T_r$ values also exceed the $\tilde T_\ell$ data at times, especially for the linear model. Further, there are times, around 500s for example, where the $\tilde T_r$ data also exceeds $\tilde T_\ell$ data. These discrepancies indicate two types of problem: the entry and exit temperature sensors have differing response times and accuracies, as is common in application; and the heat flux model in \eqref{eq:heat-flux} is too simplistic to capture the dependence of heat flux on velocity and geometry. (See \cite{holman2009heat} for more detailed analysis of these phenomena.) From a control-oriented perspective, this adds further weight to accommodating these slow variations -- we quantify time constants shortly in Subsection~\ref{subsec:isoNiso} via eigenvalue analysis --  through the design of the controller and to preserve the parsimony of the linear model, which captures the salient~dynamics.

%
%

\subsection{Linear isothermal 2D model} 
\begin{figure}[ht]
    \centering
\includegraphics[width=.95\columnwidth]{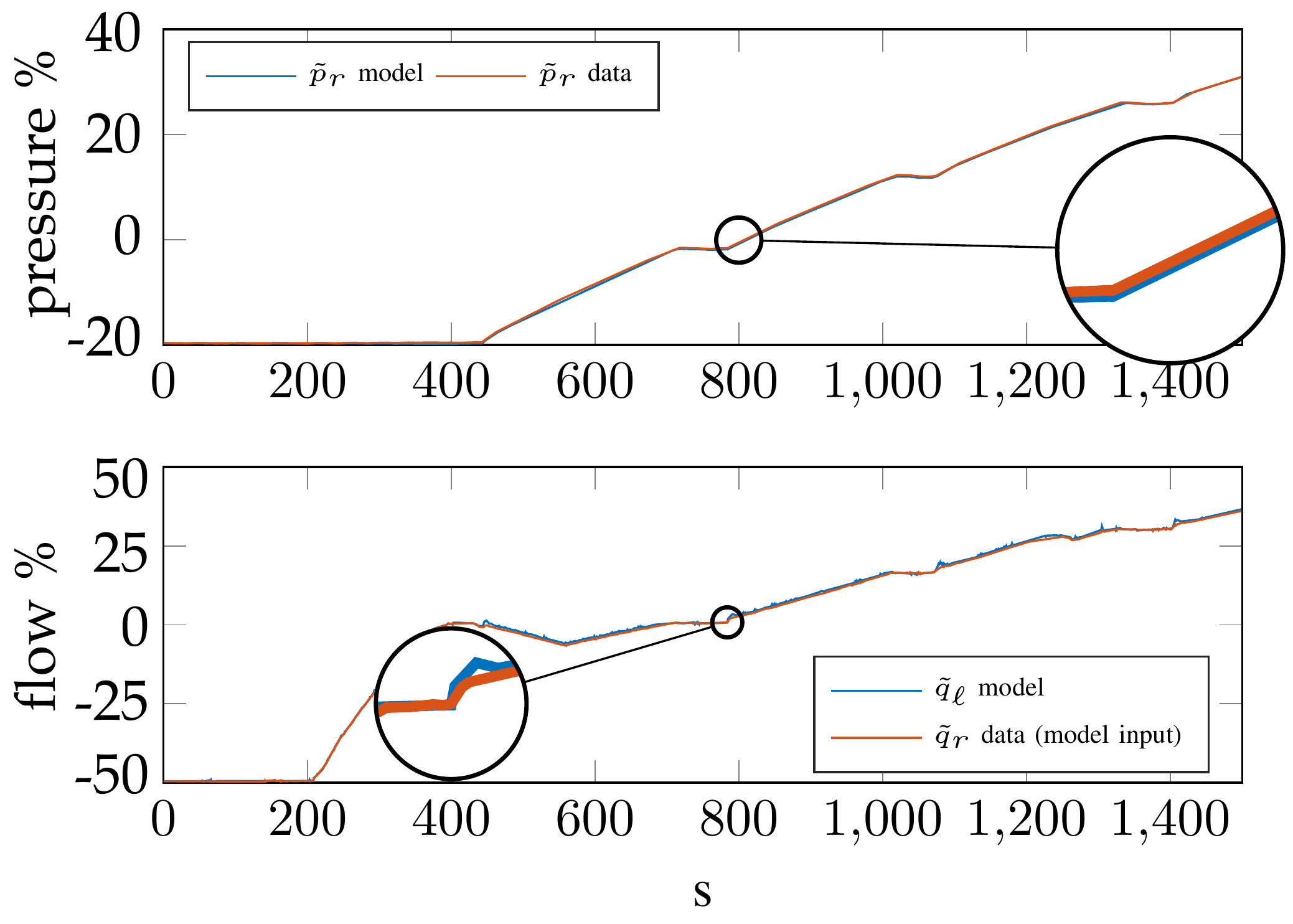}
 \caption{Isothermal 2D model with percentage deviations from the nominal point, compared against GCTF data and driven by the respective data inputs.}
    \label{fig:2d_iso_vs_data}
\end{figure}

Consider now the isothermal 2D model for which the model parameters and nominal point are equal to those of the nonisothermal 3D model above, except the temperature, which we set to $T_0=(T_{\ell,ss}+T_{r,ss})/2$. As before, $\tilde p_\ell$ and $\tilde q_r$ from the data set are model inputs, and we compare the modeled pressure $\tilde p_r$ against $\tilde p_r$ from the data. The result is shown in Figure \ref{fig:2d_iso_vs_data}. 
Notice that the modeled responses for pressure and mass flow seem congruent with those of the nonisothermal 3D model, i.e., the modeled pressure is close to the measured pressure, but displays a small static offset. The mass flows at both ends of the pipe are close, consistent with conservation of mass at steady state.

\subsection{Isothermal 2D vs. nonisothermal 3D model}\label{subsec:isoNiso}
\begin{figure}[ht]
    \centering
    \setlength\fheight{2.5cm} 
    \setlength\fwidth{.85\columnwidth}
\includegraphics{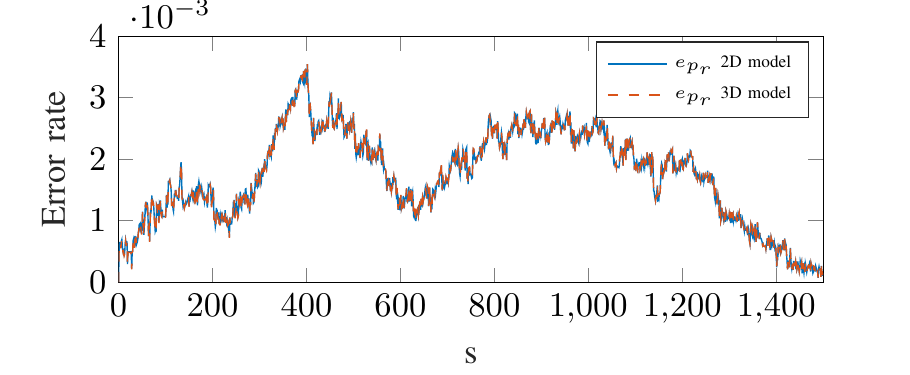} 
\caption{Respective $p_r$ pressure percentage errors of the isothermal 2D and nonisothermal 3D models.}
    \label{fig:3d_vs_2d_vs_data}
\end{figure}

The results above are now evaluated in view of the control-oriented aspect of our approach. The similarity of both the isothermal and nonisothermal model and their accuracy characterize Figure \ref{fig:3d_vs_2d_vs_data}, which shows the relative error between the modeled and measured pressure. The errors of the respective models are closely aligned, rather constant and at most at a rate of $4\times 10^{-3}$.
Both the isothermal 2D and the nonisothermal 3D linear models exhibit almost identical small offsets in simulated pressure and both capture the pressure dynamics accurately. From a control design perspective, the controller can be constructed to accommodate this modeling error.

Computing the eigenvalues of the system matrices of the related isothermal 2D and nonisothermal 3D linear models, respectively $A_{iso}$ and $A_{niso}$, and of the truncation of $A_{niso}$ to its first two rows and columns, $\left[A_{niso}\right]_{1:2}$, we have
\begin{align*}
\eig(A_{iso})&=(-3.90 \pm12.47i),\\
\eig(A_{niso})&=(-3.90 \pm 14.31i, -0.12),\\
\eig\left(\left[A_{niso}\right]_{1:2}\right)&=(-3.88 \pm14.31i)\\
&\approx \eig(A_{iso}).
\end{align*}
From this, we conclude that the temperature state is both effectively decoupled from the pressure and mass flow states and, further, governed by a time constant approximately thirty times that of the reduced-order 2D system, which preserves the dominant lightly damped oscillatory dynamics.
Consequently, for moderate temperature gradients, it is reasonable to take the temperature as a constant and employ the isothermal 2D model.

Pressing on with this control-oriented analysis, we note the respective DC gains,
\begin{align*}
\lim_{t\to\infty}x_t^\text{2D}&=-A_{iso}^{-1}B_{iso}=\begin{bmatrix}
1.004 & -600.19\\
         0    & 1
\end{bmatrix},\\
\lim_{t\to\infty}x_t^\text{3D}&=-A_{niso}^{-1}B_{niso}=\begin{bmatrix}
1.004 &-600.33 & -25.35\\
  0    &1&    0\\
   0    &0.03    &0.92
\end{bmatrix}.
\end{align*}
Continuing the discussion in Section~\ref{sec:isothermal2d}, steady-state conservation of mass flow follows for both models as the DC gains from $(\tilde p_\ell,\tilde T_\ell)\to \tilde q_\ell$ are zero and $\tilde q_r\to \tilde q_\ell$ is precisely one. For the steady-state pressure, there is (to two decimal places) a unity gain from $\tilde p_\ell\to \tilde p_r$, indicating that changes due to friction and height differences are marginal (cf. $\DqCoeffpl$ in Section \ref{sec:isothermal2d}), and a drop of similar size for both models from $\tilde q_r\to \tilde p_r$ due to additional friction (cf. $\DqCoeffql$ in Section \ref{sec:isothermal2d}) for this example. The negative gain from $\tilde T_\ell\to \tilde p_r$ for the nonisothermal model may be due to larger heat losses to the environment; a characteristic not captured by the isothermal model. Yet, given the magnitude of the SI units used here and low temperature variations in pipe elements, the consequential discrepancy is  small, as corroborated by the simulations.

The isothermal 2D model, which relies only on mass flow and pressure measurements, dovetails with the fact that especially pressure sensors (in contrast to temperature sensors) are usually well-distributed in gas processing facilities, fast and reliable. The 2D isothermal model will be used for pipe segments and the control design will be expected to accommodate the small offsets and slow variation of dynamics with changing temperatures. {The experimental results and {customary} practice {of sparse temperature measurements} (also due to {slow} temperature sensing {responses}) suggest that temperatures in typical pipes need not directly be modeled using the 3D model; exceptions are heat exchangers, compressors and other strongly temperature-affecting devices.} Driven by this evaluation, we continue the exposition with a focus on this isothermal model.

\section*{Part 2: Control-Oriented Models of Pipe Networks}

\section{DAEs, Signal Flow Graphs and Bond Graphs}
Bond graphs \cite{borutzky2010bond} provide a systematic method for deriving dynamic equations for interconnected electro-mechanical-hydraulic systems. They combine \textit{effort} variables and \textit{flow} variables, with component properties linking the two types and conservation laws and continuity governing the flows at interconnection. In the framework of fluid flow in pipe networks \cite{Lurie2008,benner2019}, this leads to a set of PDEs for the dynamics combined with algebraic equations for the constraints. Discretizing the spatial derivative yields DAE system models, which are problematic for direct control design for these interconnected systems. By contrast, Signal Flow Graphs (SFGs) correspond to systems described exclusively by ODEs; transfer functions in the linear case. Interconnected systems are directly managed by methods such as Mason's Gain Formula for the linear case, or by writing the composite state variable ODEs without algebraic constraints. It is these latter model forms, which are amenable to control design tools.

We consider three fundamental interconnections of pipe elements: series connection, branching and joining. Using the isothermal 2D model above, we develop a catalog of composite models that describe common units in the form of interconnections of pipes. In this way, algebraic constraints and DAEs will be avoided, as exemplified through the component of joining pipes introduced first. For clarity,
\begin{enumerate*}
\item we limit this section to the 2D model, but the methodology is equivalently applicable to the 3D model; and
\item without loss of generality, we assume that the steady state mass flow, $q_{ss}$, is positive.
\end{enumerate*} 
That is, $\cdot_\ell$ denotes the side where the steady state mass flow enters the pipe and $\cdot_r$ the side with an outgoing mass flow; hence the denomination \textit{joint} and \textit{branch} to come.

The reduced state vector demonstrates that an interconnection of single pipes into more complex components, with corresponding algebraic constraints, cannot immediately be translated to a SFG using only single pipe models. 
We also point out that on the contrary, bond graphs \cite{borutzky2010bond} are able to represent more complex components including algebraic constraints. However, constraints, such as those in \eqref{eq:algConJ2}, would lead to a causal conflict of type 1 and degree 1 \cite[Definition 4.19]{borutzky2010bond}, which in turn implies the existence of DAEs and therefore disaccords with our objective of control-oriented modeling.

\section{From DAEs of index 1 to composite models}\label{sec:DAE2composite}
%
%
%
%
\begin{figure}[ht]
    \centering
        \begin{subfigure}[t]{0.45\columnwidth}
    \centering
   \resizebox{\columnwidth}{!}{\includegraphics{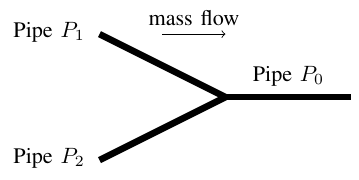}}
 \caption{Joint}
    \label{fig:joint_sketch}
    \end{subfigure}
        \begin{subfigure}[t]{0.45\columnwidth}
    \centering
   \resizebox{\columnwidth}{!}{\includegraphics{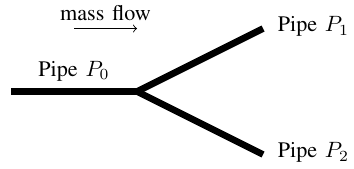}}
 \caption{Branch}
    \label{fig:branch_sketch}
    \end{subfigure}
    \caption{Pipe junctions}
    \label{fig:jb-sketch}
\end{figure}
\subsection{Joint}
Consider the joint shown in Figure~\ref{fig:joint_sketch} and let $p_{i,\ell}$ $(p_{i,r})$ be the pressure $p_\ell$ ($p_r$) of pipe $P_i$. The mass flow is denoted accordingly, so that the interconnection dictates the simplified algebraic constraints,
\begin{subequations}\label{eq:algConJ2}
\begin{align}
\tilde p_{1,r}&=\tilde p_{2,r}=\tilde p_{0,\ell},\label{eq:algConJ2_p}\\
\tilde q_{0,\ell}&=\tilde q_{1,r}+\tilde q_{2,r}.\label{eq:algConJ2_q}
\end{align}
\end{subequations}
The first equation is related to continuity and the second represents conservation of mass at the junction. This composite joint model would have a state of dimension six: $\begin{bmatrix}\tilde p_{0,r}&\tilde p_{1,r}&\tilde p_{2,r}&\tilde q_{0,\ell}&\tilde q_{1,\ell}&\tilde q_{2,\ell}\end{bmatrix}^\top$, in lexicographic ordering, plus the algebraic constraints, \eqref{eq:algConJ2}. {However, due to \eqref{eq:algConJ2_p} we can omit $\tilde p_{2,r}$ as a state (which would naturally arise in three pipe models \eqref{eq:linearPipeModel_ss_iso})}.

Define $\alpha_1$ and $\alpha_2$ to be the parameters in \eqref{eq:linearPipeModel_ss_iso} for pipes~1 and 2, and 
\begin{align}\label{eq:deltadef}
\delta=\frac{\DpCoeff[1]}{\DpCoeff[1]+\DpCoeff[2]}.
\end{align} 
Then, the six-state composite join{t} system plus constraint \eqref{eq:algConJ2} may be rewritten as an unconstrained {five}-state system
\begin{align*}
\dot x_t&=A_jx_t+B_ju_t,\\
y_t&=C_jx_t+D_ju_t,
\end{align*}
with
\begin{subequations}\label{eq:ss_junction}
\begin{align}
A_j&=
\begin{bmatrix}
0&0&
-\DpCoeff[0]&0&0\\
0&0&\DpCoeff[1](1-\delta) & -\DpCoeff[1](1-\delta)&-\DpCoeff[1](1-\delta)\\
\DqCoeffpr[0]&\DqCoeffpl[0]&\DqCoeffql[0]&0&0\\
0&\DqCoeffpr[1]&0&\DqCoeffql[1]&0\\
0  & \DqCoeffpr[2]&0&0&\DqCoeffql[2]
\end{bmatrix},\label{eq:A_j}\\
B_j&=
\begin{bmatrix}
0 & 0 & \DpCoeff[0]\\
0 & 0 & 0\\
0 & 0 & 0\\
\DqCoeffpl[1] & 0 & 0\\
0 & \DqCoeffpl[2] & 0
\end{bmatrix},
C_j=\begin{bmatrix}
1 & 0 & 0 & 0 & 0\\
0 & 0 & 0 & 1 & 0\\
0 & 0 & 0 & 0 & 1
\end{bmatrix},\\ 
D_j&=0_{3\times 3}.
\end{align}
\end{subequations}
The state, input and output vectors are now
\begin{align*}
x_t&=\begin{bmatrix}
\tilde p_{0,r} & \tilde p_{1,r} & \tilde q_{0,\ell}&\tilde q_{1,\ell}&\tilde q_{2,\ell}
\end{bmatrix}^\top,\\
u_t&=\begin{bmatrix}
\tilde p_{1,\ell} & \tilde p_{2,\ell} & \tilde q_{0,r}
\end{bmatrix}^\top,\\
y_t&=\begin{bmatrix}
\tilde p_{0,r}&\tilde q_{1,\ell}&\tilde q_{2,\ell}
\end{bmatrix}^\top.
\end{align*} 

Calculation of the steady-state gain from input three, $q_{0,r},$ to outputs two, $q_{1,\ell}$, and three, $q_{2,\ell},$ shows that, in steady state,
\begin{align*}
\tilde q_{1,\ell}+\tilde q_{2,\ell}&=\frac{\DqCoeffpr[1]\DqCoeffql[2]}{\DqCoeffpr[1]\DqCoeffql[2]+\DqCoeffpr[2]\DqCoeffql[1]}\tilde q_{0,r}
+\frac{\DqCoeffpr[2]\DqCoeffql[1]}{\DqCoeffpr[1]\DqCoeffql[2]+\DqCoeffpr[2]\DqCoeffql[1]}\tilde q_{0,r},\\
&=\tilde q_{0,r}.
\end{align*}
That is, this five-state composite join{t} model satisfies the conservation of mass flow, \eqref{eq:algConJ2_q}. Constraint \eqref{eq:algConJ2_p} is redundant, since the variables $\tilde p_{2,r}$ and $\tilde p_{0,\ell}$ have been removed; they can be computed from \eqref{eq:algConJ2_p}. The new model parameter $\delta,$ defined in \eqref{eq:deltadef}, describes the nominal proportion of flow $\tilde q_{0,\ell}$ attributed to each of the feeding pipes. This is the formal process of removing the constraint from the DAE of index 1. 

\subsection{Branch}
Differently from the joint, for  the branch in Figure~\ref{fig:branch_sketch} the equality constraint on the pressures relates the state variable, $p_{0,r}$, to input signals of the single pipe model of the branching pipes, $\tilde p_{1,\ell}$ and $\tilde p_{2,\ell}$, i.e.,
\begin{subequations}\label{eq:constraints_branch}
\begin{align}
\tilde p_{0,r}&=\tilde p_{1,\ell}=\tilde p_{2,\ell},\label{eq:c_branch_p}\\
\tilde q_{0,r}&=\tilde q_{1,\ell}+\tilde q_{2,\ell},\label{eq:c_branch_q}
\end{align}
\end{subequations}
so that the dimension of the composite model does not reduce, but is equal to the direct sum of those of the single pipe models of the individual pipes. Similarly, constraint \eqref{eq:c_branch_q} on the mass flows does not prescribe any interdependence of any input variables, but rather sets the input signal of the single pipe model of pipe $P_0$ as the sum of two other state variables. Hence, an additional parameter, such as $\delta$ for the joint is absent. The related matrices for a branch model are
\begin{subequations}\label{eq:ss_branch}
\begin{align}
A_b&=
\begin{bmatrix}0&0&0&
-\DpCoeff[0]& \DpCoeff[0] & \DpCoeff[0]\\
0&0&0&0 & -\DpCoeff[1] & 0\\
0&0&0&0 & 0 & -\DpCoeff[2]\\
\DqCoeffpr[0] & 0 & 0&\DqCoeffql[0] &0&0\\
\DqCoeffpl[1] & \DqCoeffpr[1] & 0&0&\DqCoeffql[1]&0 \\
\DqCoeffpl[2] & 0 & \DqCoeffpr[2]&0&0&\DqCoeffql[2] 
\end{bmatrix},\\
B_b&=\begin{bmatrix}
0 & 0 & 0\\
0 & \DpCoeff[1] & 0\\
0 & 0 & \DpCoeff[2]\\
\DqCoeffpl[0] & 0 & 0\\
0 & 0 & 0\\
0 & 0 & 0
\end{bmatrix},
C_b=\begin{bmatrix}
0 & 1 & 0 & 0 & 0&0\\
0 & 0 & 1 & 0 & 0&0\\
0 & 0 & 0 & 1& 0&0
\end{bmatrix},
\end{align}
\end{subequations}
with state $x_t=\begin{bmatrix}
\tilde p_{0,r} & \tilde p_{1,r} & \tilde p_{2,r} & \tilde q_{0,\ell}&\tilde q_{1,\ell}&\tilde q_{2,\ell}
\end{bmatrix}^\top$, input $u_t=\begin{bmatrix}
\tilde p_{0,\ell} & \tilde q_{1,r} & \tilde q_{2,r}
\end{bmatrix}^\top$ and output $y_t=\begin{bmatrix}
\tilde p_{1,r}&\tilde p_{2,r}&\tilde q_{0,\ell}
\end{bmatrix}^\top.$ The feedthrough matrix $D_b$ is zero.

\begin{rem}
It is straightforward to expand these ideas to intersections comprising $m$-input pipes and $n$-output pipes. This construction is available at \cite{bobSvenStarJunction} and generalizes the systematic reduction of index-1 DAEs to systems of ODEs.
\end{rem}

\subsection{Pipes in series}\label{subsec:pipes_series}
\begin{figure}[ht!]
    \centering
   \resizebox{0.9\columnwidth}{!}{\includegraphics{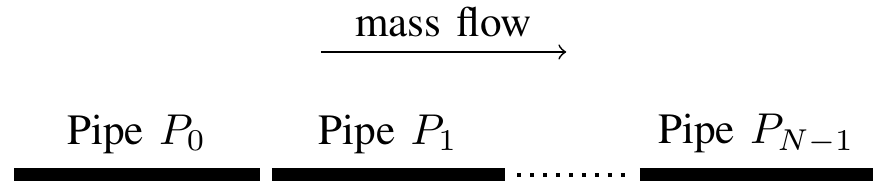}}
 \caption{Pipe series}
    \label{fig:pipe_series}
\end{figure}

$N$ pipes in series are depicted in Figure \ref{fig:pipe_series}, and are of particular interest if pipe parameters (see Table~\ref{tab:parameters}) change along the dimension of $x$ or the discretization error grows too large for a given length.
For conciseness we only state the relevant matrices here that result from the continuity conditions and conservation of mass, i.e.,
\begin{align}\label{eq:multpipe_AC}
\tilde p_{i,r}=\tilde p_{i+1,\ell}, \quad \tilde q_{i,r}=\tilde q_{i+1,\ell},
\end{align}
with $i\in\{0,1,\dots,N-2\}$.
The state, input and output elements $p_{i,r}$ and $q_{i,\ell}$ in lexicographical order, i.e., $x_t=\begin{bmatrix}
\tilde p_{0,r} & \dots & \tilde p_{N-1,r} & \tilde q_{0,\ell}&\dots & \tilde q_{N-1,\ell}
\end{bmatrix}^\top$, $u_t=\begin{bmatrix}
\tilde p_{0,\ell} & \tilde q_{N-1,r}
\end{bmatrix}^\top$ and $y_t=\begin{bmatrix}
\tilde p_{N-1,r}&\tilde q_{0,\ell}
\end{bmatrix}^\top,$
yield
\begin{align*}
A_s&=\begin{bmatrix}
\zero & A_{s,12}\\
A_{s,21} & A_{s,22}
\end{bmatrix},B_s=\begin{bmatrix}
B_{s,1}^\top & B_{s,2}^\top
\end{bmatrix}^\top,\\
C_s&=\begin{bmatrix}
\zero_{2,2(N-1)}&\eye_2&\zero_{2,2(N-1)}
\end{bmatrix},
\end{align*}
and $D_s=\zero$, where the subscripts for $\zero$ and $\eye$ describe the dimension and
\begin{align*}
A_{s,12}&=\begin{bmatrix}
-\DpCoeff[0] & \DpCoeff[0] &&\zero\\
&\ddots & \ddots&\\
&& \ddots&\DpCoeff[N-2]\\
\zero&& &-\DpCoeff[N-1]\\
\end{bmatrix},\\
A_{s,21}&=\begin{bmatrix}
\DqCoeffpr[0] &&&\zero\\
\DqCoeffpl[1]&\ddots & &\\
&\ddots& \ddots&\\
\zero&& \DqCoeffpl[N-1]&\DqCoeffpr[N-1]\\
\end{bmatrix},\\
A_{s,22}&=\diag(\DqCoeffql[0], \DqCoeffql[1], \dots, \DqCoeffql[N-1]),\\
B_{s,1}&=\begin{bmatrix}
\zero_{2(N-1),2} \\ \begin{matrix}
0 & \DpCoeff[N-1]
\end{matrix}
\end{bmatrix},B_{s,2}=\begin{bmatrix}
\begin{matrix}
\DqCoeffpl[0] & 0
\end{matrix}\\ \zero_{2(N-1),2}
\end{bmatrix}.
\end{align*} 
Since each pipe conforms to steady-state conservation of mass flow, the interconnection automatically does {as well}. Bode diagrams are provided in \cite{bobSvenStarJunction} for 2i2o models of a 30-meter pipe section as: a single 30m pipe, two 15m pipes in series, three 10m pipes in series; low-frequency responses coincide.

\section{Systematic Model Interconnection}\label{sec:connections}
Building on the composite models above, we introduce a matrix formulation that enables the construction of state-space models for interconnected components of pipes, joints and branches.

\subsection{Matrix methodology} Towards this goal, with $N$ interconnected components, let $\U_i\subset \R^{n_{u,i}}$ and $\Y_i\subset\R^{n_{y,i}}$ be the set of inputs and outputs for component $i$, respectively, with state $x^{(i)}_t\in\R^{n_{x,i}}$, input $u_t^{(i)}\in\R^{n_{u,i}}$ and output $y_t^{(i)}\in\R^{n_{y,i}}$. The model matrices are denoted accordingly. Further, $\overline \U$ and $\overline \Y$ are the respective sets of external inputs and outputs.
We assume that the input of every component is either connected to the output of another component or represents an external input. 
\begin{ass}[connectedness]\label{ass:inputOutputConnected}
Let $i\in \{1,2,\dots,N\}$. For any $u\in\U_i$ there exists a $j\in \{1,2,\dots,N\}, j\neq i,$ such that $u\in\Y_j\cup \overline \U$.
\end{ass}
Internal interconnections are governed by the continuity of pressure and mass flow per \eqref{eq:multpipe_AC}. We begin by stacking the state-space models of the individual network components. {With some abuse of notation,}
\begin{align}\label{eq:totalSystem}
\dot{\bar x}_t=A\bar x_t+B\bar w_t,\quad \bar y_t=C\bar x_t+D\bar w_t
\end{align}
where
\begin{subequations}\label{eq:matrices_stacked_up}
\begin{align}
A&=\blkdiag(A^{(1)}, A^{(2)}, \dots, A^{(N)}),\\
B&=\blkdiag(B^{(1)}, B^{(2)}, \dots, B^{(N)}),\\
C&=\blkdiag(C^{(1)}, C^{(2)}, \dots, C^{(N)}),\\
D&=\blkdiag(D^{(1)}, D^{(2)}, \dots, D^{(N)}).
\end{align}
\end{subequations}
The total state vector, $\bar x_t\in\R^{n_{x,1}+n_{x,2}+\dots+n_{x,N}}$, total output vector, $\bar y_t\in\R^{n_{y,1}+n_{y,2}+\dots+n_{y,N}}$, and \emph{component} input vector, $\bar w_t\in\R^{n_{u,1}+n_{u,2}+\dots+n_{u,N}}$ are composed correspondingly of direct sums. Assumption~\ref{ass:inputOutputConnected} yields
\begin{align}\label{eq:totalInput}
\bar w_t=F\bar y_t+G\bar u_t,
\end{align}
where $\bar u_t$ denotes the total input vector of external signals feeding into the total system. The matrix $F$ describes connections between component inputs and outputs, and $G$ is related to the connection between external inputs and internal components. Hence, both matrices are sparse and constructed as follows,
\begin{subequations}\label{eq:connection_matrices}
\begin{align}
[F]_{i,j}&=\begin{cases}
1, & \text{if } [\bar y_t]_j=[\bar w_t]_i,\\
0, & \text{otherwise},
\end{cases}\\
[G]_{i,j}&=\begin{cases}
1, & \text{if } [\bar u_t]_j=[\bar w_t]_i,\\
0, & \text{otherwise},
\end{cases}
\end{align}
\end{subequations}
with $[\cdot]_{i,j}$ denoting the matrix element in row $i$ and column $j$, and vectors written accordingly. This formulation is applied to a numerical example in Section \ref{sec:num_ex}. It allows us to define a state model of the total system.
\begin{prop}\label{prop:overallStateSpace}
A (perhaps non-minimal) state-space realization of \eqref{eq:totalSystem} and \eqref{eq:totalInput} is given by
\begin{align}\label{eq:matrix_framework}
\bar x_t=\bar A \bar x_t+\bar B \bar u_t, \quad \bar y_t=\bar C \bar x_t+\bar D \bar u_t,
\end{align}
where
\begin{align*}
\bar A&=A+BF(I-DF)^{-1}C, \quad \bar C=(I-DF)^{-1}C,\\
\bar B&=B\left[I+F(I-DF)^{-1}D\right]G,\quad \bar D=(I-DF)^{-1}DG.
\end{align*}
\end{prop}
\begin{pf}
Substituting $\bar w$ from \eqref{eq:totalInput} into \eqref{eq:totalSystem} for $\bar y_t$ yields
\begin{align}
\bar y&=(I-DF)^{-1}C\bar x_t+(I-DF)^{-1}DG\bar u_t,\nonumber\\
\bar w&=F(I-DF)^{-1}C\bar x_t+\left[I+F(I-DF)^{-1}D\right]G\bar u_t.\label{eq:weqn}
\end{align}
Now substituting from \eqref{eq:weqn} into \eqref{eq:totalSystem} for $\dot {\bar x}_t$ produces the closed-loop state-space connected system.
\begin{align*}
\dot {\bar x}_t&=\left[A+BF(I-DF)^{-1}C\right]\bar x_t\nonumber \\
&\quad+B\left[I+F(I-DF)^{-1}D\right]G\bar u_t,\\
\bar y_t&=(I-DF)^{-1}C\bar x_t+(I-DF)^{-1}DG\bar u_t.
\end{align*}
\end{pf}
Here, the total output, $\bar y_t$, is set to be the outputs of all components. However, if only some variables constitute to the total output modifying $\bar y_t$ is a simple exercise through the multiplication of $\bar C$ and $\bar D$ by an appropriate selection matrix.
\subsection{Subsuming Mason}
Next, we show that the state-space realization above subsumes Mason's Gain Formula \cite{mason}. The latter is a method to find transfer functions of SFGs with multiple inputs and multiple outputs and has also been established in a simple matrix form in e.g. \cite{HuaichenChenMasonChineseJelectronics2002}. The interest in this equivalence result lies in its generality for linear systems and advantage over Mason's Gain Formula via simple matrix manipulation without relying on symbolic matrix inversions with transfer functions as matrix elements. Further, the  calculation in Proposition~\ref{prop:overallStateSpace} yields all the closed-loop transfer functions between each input and each output, versus Mason, which computes SISO transfer functions using Cramer's Rule.

Mason's Gain {Formula} formulation in \cite{HuaichenChenMasonChineseJelectronics2002} starts by writing the vector of output signals, $\bar y$, as the interconnection of $y_t^{(i)}$ and $\bar u_t$ with transfer function matrices,
\begin{align}
\bar y_t&=\mathcal Q\bar y_t+\mathcal P\bar u_t.\label{eqLpremason}
\end{align} 
Mason's Gain {Formula} is then that the solution is given by
\begin{align}
\bar y_t&=(I-\mathcal Q)^{-1}\mathcal P\bar u_t.\label{eq:mason}
\end{align}
\begin{prop}\label{prop:mason}
A state-variable realization of Mason's Gain {Formula} transfer function, $(I-\mathcal Q)^{-1}\mathcal P,$ is given in \eqref{eq:matrix_framework}.
\end{prop}
This is proven in the Appendix.

\section{Numerical experiment}\label{sec:num_ex}
\begin{figure}[ht!]
    \centering
   \includegraphics{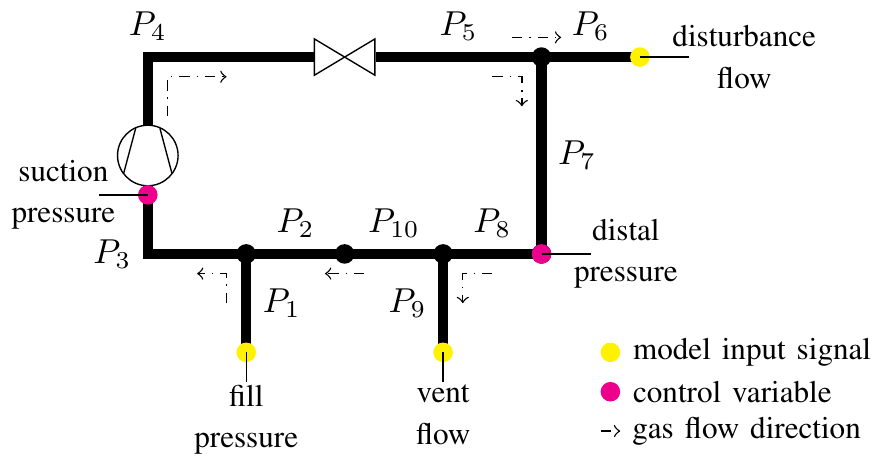}
  \caption{Pipe network with compressor and valve $\bowtie$. In process control parlance, the fill pressure and vent flow are manipulated variables, the suction and distal pressures are controlled variables, and the flow from $P_6$ is a disturbance signal.}
    \label{fig:control_loop}
\end{figure}

We apply our modeling methodology to the loop illustrated in Figure \ref{fig:control_loop}, which represents a hypothetical pipe loop at the GCTF. Such a feedback system creates problems for  DAE methods, such as those in \cite{benner2019}, because of the algebraic constraints. Here we use it as a proof-of-concept test case and rely, rather unrealistically but similarly to \cite{benner2019} for distribution networks, on isothermal models and treatment of the compressor and valve as static gains. Clearly, the thermal properties of compressors, heat exchangers and valves play an important role on the spatial scales of gas processing facilities and these will form the focus for ongoing modeling.

%
%
The gas is methane and flows clockwise, entering through pipe $P_1$ and exiting through pipes $P_6$  and $P_9$. The aim is to regulate the pressures $p_{3,r}$ and $p_{7,r}$ in the face of leakage via $P_6.$ The Haaland formula \cite{rennels2012pipe} and assumed parameters\footnote{\label{foot:params} All pipes are assumed to have the same geometry, i.e., $L=10\text{m}, d=0.7\text{m}, \text{roughness}=4.57\times 10^{-5}\text{m}$. Further, we assume that $Re\approx 1.168\times10^{8}, T_0=300\text{K}, z_0=0.95,R_s=518.28\text{J}/\text{(Kmol)}$, $\tilde p_{ss,\ell}=25\times 10^5\text{Pa}$ and $\tilde q_{ss}=21\text{m}/\text{s}^2$.} yield a friction factor for each pipe of $\lambda=0.0111$.
\subsection{Network model} The compressor and valve, whose corresponding variables are respectively labeled by subscripts $c$ and $v$, are modeled as static gains
\begin{align*}
D_c=\begin{bmatrix}
k_c&0\\0 & 1
\end{bmatrix},\quad D_v=\begin{bmatrix}
k_v&0\\0 & 1
\end{bmatrix},
\end{align*}
where $k_c=4$ and $k_v=0.8$. Further, pipes $(P_1,P_2,P_3)$ are modeled as a joint, as in \eqref{eq:ss_junction}, and $(P_5,P_6,P_7)$ and $(P_8,P_9,P_{10})$ as branches, as in \eqref{eq:ss_branch}.
Composing the system according to \eqref{eq:matrices_stacked_up}, results in the component input vector,
\begin{align*}
\bar w_t&=\left[\begin{matrix}
\tilde p_{1,\ell}& \tilde p_{2,\ell} & \tilde q_{3,r} & \tilde p_{c,\ell} & \tilde q_{c,r} & \tilde p_{4,\ell} & \tilde q_{4,r} \end{matrix}\right.\\
&\qquad \left.\begin{matrix} \tilde p_{v,\ell} & \tilde q_{v,r} & \tilde p_{5,\ell} & \tilde q_{6,r} & \tilde q_{7,r} & \tilde p_{8,\ell}&\tilde q_{9,r} & \tilde q_{10,r}
\end{matrix}\right]^\top,
\end{align*}
and the total output vector,
\begin{align*}
\bar y_t&=\left[\begin{matrix}
\tilde p_{3,r}& \tilde q_{1,\ell} & \tilde q_{2,\ell} & \tilde p_{c,r} & \tilde q_{c,\ell} & \tilde p_{4,r} & \tilde q_{4,\ell} \end{matrix}\right.\\
&\qquad \left.\begin{matrix} \tilde p_{v,r} & \tilde q_{v,\ell} & \tilde p_{6,r} & \tilde p_{7,r} & \tilde q_{5,\ell} & \tilde p_{9,r}&\tilde p_{10,r} & \tilde q_{8,\ell}
\end{matrix}\right]^\top.
\end{align*}
The inputs of the total system are
\begin{align*}
\bar u_t&=\begin{bmatrix}
\tilde p_{1,\ell} & \tilde q_{6,r} & \tilde q_{9,r}
\end{bmatrix}^\top.
\end{align*}

With \eqref{eq:connection_matrices}, the total input and output vector, $\bar u_t$ and $\bar y_t$, as well as the component input vector, $\bar w_t$, are the basis for the construction of $F$ and $G$. For example, $[\bar w_t]_1=\tilde p_{1,\ell}$ is an input of the total system and the first element of $\bar u_t$. Hence $[G]_{1,1}=1$. Further, $[\bar w_t]_2=\tilde p_{2,\ell}$ connects to $\tilde p_{10,r}=[\bar y_t]_{14}$, so that $[F]_{2,14}=1$. Similarly, $[\bar w_t]_{11}=\tilde q_{6,r}$ is another total input, i.e., $\tilde q_{6,r}=[\bar u_t]_{2}$, so that $[G]_{11,2}=1$. In this way, by passing through $\bar w_t$ and following \eqref{eq:connection_matrices}, we can fill the matrices with ones at the appropriate location and zeros otherwise. {The eigenvalues of the resulting interconnected system all have negative real part; hence stability is {demonstrated}. Some eigenvalues {have} large imaginary parts pointing to the high-oscillatory resonant modes, which we ignore in the control design, which will recognize the presence of anti-aliasing filters in the sensors \cite{sven_bob_control}.}

\subsection{Steady state: conservation of mass}
The isothermal LTI closed-loop system is stable with the overall pressure static gains from $\tilde p_{1,\ell}$ to all but $\tilde p_{2,r}$ greater than one. Increasing the compressor and/or valve gains can bring about instability, as might be expected. Further, since the frequency response of each component is available, standard stability tests may be performed. Indeed, the control design is to construct a stabilizing 2-input/2-output regulator to reject the effect of the disturbance flow.

To evaluate the model in terms of conservation of mass, we also analyze the steady-state gains from the three loop inputs, $\tilde p_{1,\ell}$, $\tilde q_{6,r}$ and $\tilde q_{9,r}$, to each pipe's mass flow. The corresponding DC-gain values are shown in Table~\ref{table:flows}. Each column represents one model input and each row shows the corresponding steady-state change in mass flow from nominal due to a unit step change of the respective input and zero inputs otherwise.
\begin{table}[h]
\begin{centering}
\begin{tabular}{c|lll}
to\textbackslash from&fill: $\tilde p_{1,\ell}$&vent: $\tilde q_{9,r}$&dist: $\tilde q_{6,r}$\\
\hline
$\tilde q_{1,\ell}$&	$0$ & $1$ & $1$ \\
$\tilde q_{2,\ell}$&     $0.184$ & $-1.022$ & $-0.8$ \\
$\tilde q_{3,\ell}$&     $0.184$ & $-0.022$ & $0.2$ \\
$\tilde q_{4,\ell}$&     $0.184$ & $-0.022$ & $0.2$ \\
$\tilde q_{5,\ell}$&     $0.184$ & $-0.022$ & $0.2$ \\
$\tilde q_{6,\ell}$&     $0$ & $0$ & $1$ \\
$\tilde q_{7,\ell}$&     $0.184$ & $-0.022$ & $-0.8$ \\
$\tilde q_{8,\ell}$&     $0.184$ & $-0.022$ & $-0.8$ \\
$\tilde q_{9,\ell}$&     $0$ & $1$ & $0$ \\
$\tilde q_{10,\ell}$&     $0.184$ & $-1.022$ & $-0.8$\\
\hline
\end{tabular}
\caption{DC (steady-state) gains from inputs to mass flows.\label{table:flows}}
\end{centering}
\end{table}
\subsubsection{Step response fill pressure change} Evaluating the first column with input $\tilde p_{1,\ell}$, a zero change in mass flows $\tilde q_{6,\ell},\tilde q_{9,\ell}$ is consistent the other zero inputs, $\tilde q_{9,r}=\tilde q_{6,r}=0$. As a result, the steady-state mass flow $\tilde q_{1,\ell}=0$. A higher fill pressure leads to a larger mass flow around the loop, uniformly through all pipes, as evident by the numerical values of the other rows of the same column.
\subsubsection{Step responses vent and disturbance flow changes} Evaluating the second column with input $\tilde q_{9,r}=1$ and zero disturbance flow, i.e. $\tilde q_{6,r}=0$ (and hence $\tilde q_{6,\ell}=0$), $1\text{kg/s}^2$ enters the loop through $\tilde q_{9,r}$ so that $\tilde q_{9,\ell}=1$. We further note that mass flow around the loop uniformly dropped by $-0.022$ excluding pipes $P_{10}$ and $P_2$. The flow through Pipe $P_{10}$ and $P_2$ reduces by $-1.022$ as a result of the reduced overall flow and unit flow exiting through pipe $P_9$. Then, the additional flow $\tilde q_{1,\ell}=1$ through pipe $P_1$ brings the flow back to $-0.022$. The same reasoning can be applied to the last column related to the disturbance input $\tilde q_{6,r}$.

Our analysis shows that conservation of mass around the loop is captured through the use of composite models and the matrix methodology presented above, without imposing additional algebraic constraints. Further, the linear time-invariant model is amenable to direct feedback controller design and stability analysis.

\section{Conclusion and further directions}
In this paper, we present control-oriented models in the form of LTI state-space realizations that capture the dominant dynamics for the pressure, mass flow and temperature in pipes at a scale appropriate for gas processing facilities. Validation against real-world data and simulation of the initial constituent equations illustrate their suitability for model-based controller design, which will incorporate requirements for robustness to minor static offsets and slow variations. Building on these models, we elaborate on the need for composite elements for interconnections to absorb DAEs, and provide a corresponding catalog of composite models for common units. To increase practical relevance of the proposed model, we also introduce a matrix methodology that enables a simple creation of pipe networks and illustrate its behavior with a numerical experiment. The analysis of costs and benefits of nonisothermal models indicates and quantifies inaccuracies of the models and distinguishes between models parametrized by nominal temperature versus those parametrized by measured temperatures. {Here, we focus on process control; additional (nonlinear) control systems across multiple operating points may be employed for safety, start-up and shutdown and these could be local to specific units, and rapid in their action. Our methods are not targeted towards these controllers.}

{The control-oriented modeling developed here draws guidance at the formulation stage from the control objective specification in the introduction. Two companion works take these methods further. In \cite{SRB_arXiv}, the modeling methods are applied to a generate linear state-space models for a wider variety of network elements. This paper provides a compendium of modeled elements together with their derivation and proof of {internal} satisfaction of conservation rules. The compendium also provides example \matlab\ code illustrating the connection process for models. Thus, \cite{SRB_arXiv} is a support document. The technical partner paper \cite{sven_bob_control} on the other hand marries the control-oriented modeling with model-based control  and provides strong evidence of the role played by model features here in subsequent controller development. Particularly, {\cite{sven_bob_control}} explores in detail the regulation control effect of the {mass-conserving} models.}

\section*{Acknowledgement}
This research was supported by funding from Solar Turbines {Incorporated}, who also provided operating data and guidance.

\bibliographystyle{plain}
\bibliography{/Users/sven/Documents/MEGA/Uni/Latex_ressources/bib_all.bib}

\appendix
\section{Proofs}
\subsubsection*{Proposition \ref{prop:3D_pde}} \hfill\\
By hypothesis and with $ q=\rho A_{c}v$, equation \eqref{eq:pde_q_noniso} results directly from the Momentum and Gas Equations, \eqref{eq:momentum} and \eqref{eq:ideal-gas}, see e.g. \cite{benner2019}.

For the pressure-related PDE, let $a_1\doteq \frac{c_v}{R_sz_0}-\frac{R_sz_0}{2A_{c}^2}\frac{q^2 T}{ p^2}+\frac{gh}{R_s Tz_0}$. Then, solving energy equation \eqref{eq:energy} for $\frac{\partial p}{\partial t}$ yields
\begin{align*}
\frac{\partial  p}{\partial t}&=
a_1^{-1}\left(\q\rho +\frac{\partial  T}{\partial t}\underbrace{\left(\frac{gh p}{R_s T^2z_0}- \frac{R_s z_0 q^2}{2A_{c}^2 p}\right)}_{\doteq -a_2}
-\frac{R_sz_0 T q}{A_{c}^2 p}\frac{\partial  q}{\partial t}\right.\\
& \qquad \left.\vphantom{\underbrace{\frac{a^2}{a}}_{a_2}}-\frac{\partial}{\partial x}\left[\frac{ q}{A_{c}}\left(c_v T+gh\right)+\frac{ q^3R_s^2 T^2z_0^2}{2 p^2A_{c}^3}+\frac{ qR_s Tz_0}{A_{c}}\right]\right).
\end{align*}
From Continuity and Gas Equations, resp. \eqref{eq:continuity} and \eqref{eq:ideal-gas},
\begin{align}\label{eq:proof_Tdot}
\frac{\partial  T}{\partial t}=\frac{z_0R_s T^2}{ pA_{c}}\frac{\partial  q}{\partial x}+\frac{ T}{ p}\dot { p},
\end{align}
since $(a_1+\frac{ T}{ p}a_2)^{-1}=\frac{R_sz_0}{c_v}$ and, using \eqref{eq:pde_q_noniso}, yields
\begin{align*}
\frac{\partial  p}{\partial t}&=\frac{R_sz_0}{c_v} \left(\q\rho +\frac{\partial  q}{\partial x}\left(\frac{gh}{A_{c}}+ \frac{3R_s^2 z_0^2 q^2}{2A_{c}^3 p^2}\right)
+\frac{R_sz_0 T q}{A_{c}^2 p}\left[A_{c}\frac{\partial  p}{\partial x}\right.\right.\\
&\quad \left.\left.+A_{c}\left(\frac{z_0R_s q^2}{A_{c}^2 p}\frac{\partial  T}{\partial x}-\frac{z_0R_s T q^2}{A_{c}^2 p^2}\frac{\partial  p}{\partial x}\right)+\frac{\lambda z_0R_s T}{2DA_{c} p} q| q|\right.\right.\\
&\quad\left.\left.+gA_{c}\frac{ p}{z_0R_s T}\frac{dh}{d x}\right]\right.\\
&\qquad \left.-\frac{\partial}{\partial x}\left[\frac{ q}{A_{c}}\left(c_v T+gh\right)+\frac{ q^3R_s^2 T^2z_0^2}{2 p^2A_{c}^3}+\frac{ qR_s Tz_0}{A_{c}}\right]\right).
\end{align*}
Computing the spatial derivative leads to
\begin{align*}
\frac{\partial  p}{\partial t}&=\frac{R_sz_0}{c_v A_{c}}\left[\q\rho A_{c}-\frac{\partial  q}{\partial x} T\left(c_v+R_sz_0\right)+\frac{\partial  p}{\partial x} \frac{R_sz_0 T q}{ p}\right.\\
&\qquad \left.-\frac{\partial  T}{\partial x} q\left(c_v+R_sz_0\right)+\frac{\lambda R_s^2z_0^2 T^2 q^2| q|}{2DA_{c}^2 p^2}\right],
\end{align*}
which with \eqref{eq:heat-flux} results in \eqref{eq:pde_p_noniso}.

For the temperature, we use the result above and \eqref{eq:proof_Tdot} to obtain
\begin{align*}
\frac{\partial  T}{\partial t}&=\frac{R_sz_0 T}{c_v A_{c} p}\left[\q\rho A_{c}-\frac{\partial  q}{\partial x} TR_sz_0+\frac{\partial  p}{\partial x} \frac{R_sz_0 T q}{ p}\right.\\
&\qquad \left.-\frac{\partial  T}{\partial x} q\left(c_v+R_sz_0\right)+\frac{\lambda R_s^2z_0^2 T^2 q^2| q|}{2DA_{c}^2 p^2}\right],
\end{align*}
which with \eqref{eq:heat-flux} results in \eqref{eq:pde_T_noniso}.

\subsubsection*{Proposition \ref{prop:mason}} \hfill\\
For readability we exclude the subscript $t$ in this proof. From \eqref{eq:totalSystem}, state-variable realizations of the transfer functions $\mathcal Q$ and $\mathcal P$ are given by the following:
\begin{align}
\mathcal Q&=[D+C(sI-A)^{-1}B]F,\label{eq:Qdef}\\
\mathcal P&=[D+C(sI-A)^{-1}B]G.\label{eq:Pdef}
\end{align}
Firstly, use the matrix inversion formula to write
\begin{align}
(I-\mathcal Q)^{-1}&=\{I-[D+C(sI-A)^{-1}B]F\}^{-1},\nonumber\\
&=(I-DF)^{-1}+\nonumber \\
&\quad (I-DF)^{-1}C[sI-A-BF(I-DF)^{-1}C]^{-1}\nonumber\\
&\quad BF(I-DF)^{-1}.\label{eq:IQ}
\end{align}
Define the $\bar u$-component  of $\bar w$ as $\bar w^{\bar u}=[I+F(I-DF)^{-1}D]G\bar u=w^{\bar u}_1+Fw^{\bar u}_2$ with $w^{\bar u}_1=G\bar u$ and $w^{\bar u}_2=(I-DF)^{-1}DG\bar u$ and appeal to linearity to define
\begin{align*}
\dot{\bar x}_1&=[A+BF(I-DF)^{-1}C]\bar x_1+B\bar w^{\bar u}_1,\\
\dot {\bar x}_2&=[A+BF(I-DF)^{-1}C]\bar x_2+BF\bar w^{\bar u}_2,\\
\bar y&=(I-DF)^{-1}C(\bar x_1+\bar x_2)+\bar w^{\bar u}_2.
\end{align*}
From here and the definitions of $w^{\bar u}_1$ and $w^{\bar u}_2$, it is apparent that $\bar y=\bar y_1+\bar y_2$ where,
\begin{align}
\dot{\bar x}_1&=[A+BF(I-DF)^{-1}C]\bar x_1+BG\bar u,\label{eq:sys1state}\\
\bar y_1&=(I-DF)^{-1}C\bar x_1,\label{eq:sys1out}
\end{align}
and
\begin{align}
\dot{\bar x}_2&=[A+BF(I-DF)^{-1}C]\bar x_2+BF(I-DF)^{-1}DG\bar u,\label{eq:sys2state}\\
\bar y_2&=(I-DF)^{-1}C\bar x_2+(I-DF)^{-1}DG\bar u,\label{eq:sys2out}
\end{align}

\textit{{System~1} \eqref{eq:sys1state}-\eqref{eq:sys1out}}:
Denote the system transfer function $\mathcal K=(I-DF)^{-1}C(sI-A)^{-1}B$ and rewrite \eqref{eq:sys1state} as output feedback around $\mathcal K$.
\begin{align*}
\dot{\bar x}_1&=A\bar x_1+B(F\bar y_1+G\bar u).
\end{align*}
In turn, writing this in terms of $\mathcal K$ and using \eqref{eq:IQ} for $(I-\mathcal Q)^{-1},$
we have
\begin{align}
\bar y_1&=\mathcal K(F\bar y_1+G\bar u)\nonumber\\
&=(I-\mathcal KF)^{-1}\mathcal KG\bar u\nonumber\\
&=[I-(I-DF)^{-1}C(sI-A)^{-1}BF]^{-1}\nonumber \\
&\quad (I-DF)^{-1}C(sI-A)^{-1}BG\bar u\nonumber\\
&=\{I+(I-DF)^{-1}C[sI-A-BF(I-DF)^{-1}C]^{-1}\nonumber\\
&\quad BF\}(I-DF)^{-1}C(sI-A)^{-1}BG\bar u\nonumber\\
&=\{(I-DF)^{-1}+(I-DF)^{-1}C[sI-A-BF\nonumber\\
&\quad (I-DF)^{-1}C]^{-1}BF(I-DF)^{-1}\}C(sI-A)^{-1}BG\bar ,\nonumber\\
&=(I-\mathcal Q)^{-1}C(sI-A)^{-1}BG\bar u.\label{eq:parti}
\end{align}

\textit{{System~2} \eqref{eq:sys2state}-\eqref{eq:sys2out}}:
Directly comparing \eqref{eq:IQ} to \eqref{eq:sys2state}-\eqref{eq:sys2out}, we see that 
\begin{align}
\bar y_2&=(I-\mathcal Q)^{-1}\,DG\bar u.\label{eq:partii}
\end{align}

Combining \eqref{eq:parti} and \eqref{eq:partii} we have
\begin{align*}
\bar y&=\bar y_1+\bar y_2,\\
&=(I-\mathcal Q)^{-1}C(sI-A)^{-1}BG\bar u+(I-\mathcal Q)^{-1}\,DG\bar u,\\
&=(I-\mathcal Q)^{-1}\mathcal P\bar u.
\end{align*}\hfill$\blacksquare$

\end{document}